\documentclass[aps,pre,
twocolumn,
showpacs]{revtex4}

\usepackage{amsmath}
\usepackage{amssymb}
\usepackage{graphicx}
\usepackage{bm}
\usepackage{epstopdf}
\usepackage{graphicx,epstopdf}
\usepackage{amsmath,amsfonts,latexsym}

\newcommand{\wtH}{\widetilde{H}}
\newcommand{\wtU}{\widetilde{U}}
\newcommand{\sech}{\rm sech}
\newcommand{\abs}[1]{\left|#1\right|}

\begin{document}

\title {On the integrable elliptic cylindrical Kadomtsev-Petviashvili equation}

\author {K.R. Khusnutdinova$^{1}$}
\thanks{Corresponding author. E-mail: K.Khusnutdinova@lboro.ac.uk; Tel: +44 (0)1509 228202; Fax: +44 (0)1509 223969. }
\author{C. Klein$^{2}$}
\author{V.B. Matveev$^{2}$}
\author{A.O. Smirnov$^{3}$}

\address{
$^{1}$Department of Mathematical Sciences, Loughborough University,   Loughborough LE11 3TU, UK\\
$^{2}$Institut de Math\'ematiques de Bourgogne, BP 47870-F-21078, Dijon, France\\
$^{3}$ St. Petersburg University of Aerospace Instrumentation,  St.Petersburg, 190000, Russia\\
 }

\date{\today}

\
\begin{abstract}
There exist two versions of the Kadomtsev-Petviashvili equation, related to the Cartesian and cylindrical geometries of the waves. In this paper we derive and study a new version, related to the elliptic cylindrical geometry.  The derivation is given in the context of surface waves, but the derived equation is a universal integrable model applicable to generic weakly-nonlinear weakly-dispersive waves. We also show that there exist nontrivial transformations between all three versions of the KP equation associated with the physical problem formulation, and use them to obtain new classes of approximate solutions for water waves.

\end{abstract}


\maketitle

{\bf The `elliptic cylindrical Kadomtsev-Petviashvili (ecKP) equation',
\begin{eqnarray*}
&&\left(H_\tau + 6 H H_\zeta +  H_{\zeta \zeta \zeta} +  \frac{\tau}{2 (\tau^2 - a^2)} H \right. \\
&&\left. - \frac{a^2 \nu^2}{12 \sigma^2 (\tau^2 - a^2)}  H_\zeta   \right) _\zeta +  \frac{3 \sigma^2}{\tau^2-a^2} H_{\nu \nu}= 0,
\end{eqnarray*}
where $a$ is a parameter and $\sigma^2 = \pm 1$,
is derived for surface gravity waves with nearly-elliptic front, generalising the cylindrical KP equation  for nearly-concentric waves and describing the intermediate asymptotics.  We find transformations between the derived ecKP equation and two existing versions of the KP equation for water wave problems, for nearly-plane and nearly-concentric waves, as well as the Lax pair for the ecKP equation. The transformations are used to construct  important classes of exact solutions of the derived ecKP equation and corresponding new asymptotic solutions for the Euler equations  from the known solutions of the  KP equation. The ecKP equation is a universal  integrable model applicable to generic weakly-nonlinear weakly-dispersive waves with nearly-elliptic wave fronts.}

\section{Introduction}

There exist two classical versions of the Kadomtsev-Petviashvili (KP) equation \cite{KP} associated with the surface wave problems for an incompressible fluid, described by the full set of Euler equations with free surface and rigid horizontal bottom boundary conditions (see \cite{AS1,Johnson1, AS, Johnson2} and references therein):
\begin{eqnarray}
&&\rho (u_t + uu_x + vu_y + wu_z) = -p_x, \nonumber \\
&&\rho (v_t + uv_x + vv_y + wv_z) = -p_y, \nonumber \\
&&\rho (w_t + uw_x + vw_y + ww_z) = -p_z-\rho g, \nonumber \\
&&u_x + v_y + w_z = 0, \nonumber \\
&&p|_{z = h(x,y,t)} = p_a \nonumber \\
&&- \Gamma \frac{(1+h_y^2) h_{xx} + (1 + h_x^2) h_{yy} - 2 h_x h_y h_{xy}}{(1 + h_x^2 + h_y^2)^{3/2}}, \nonumber \\
&&w|_{z = h(x,y,t)} = h_t+uh_x+vh_y, \nonumber \\
&&w|_{z=0} = 0.
\label{Euler}
\end{eqnarray}
Here, $(u,v,w)$ are the three components of the velocity vector in 
Cartesian coordinates $(x,y,z)$,  $t$ is the time, $p$ the pressure 
($p_a$ is the constant atmospheric pressure at the surface, and 
$\Gamma$ is the coefficient of the surface tension), $\rho$ is 
constant density, $g$ is the gravitational acceleration, $z=0$ is the bottom, and $z = h(x,y,t)$ is the free surface.
The original KP equation \cite{KP},
\begin{equation}
\left (U_{\tau} + 6 U U_\xi + U_{\xi \xi \xi} \right )_{\xi} + 3 \sigma^2 U_{YY} = 0,
\label{KP}
\end{equation}
 and the cylindrical KP (cKP) equation \cite{Johnson1},
\begin{equation}
\left (W_{\tau} + 6 W W_\chi + W_{\chi \chi \chi} + \frac{W}{2 \tau} \right )_{\chi} + \frac{3 \sigma^2}{\tau^2} W_{VV} = 0,
\label{cKP}
\end{equation}
are derived for the leading order term of the asymptotic expansion of the free surface elevation in the appropriate sets of fast and slow variables, and describe the weakly-nonlinear evolution of long nearly-plane and nearly-concentric waves, respectively.  For surface gravity waves with no or weak surface tension
one has $\sigma^2=1 $ (KP-II), while the case $\sigma^2 = -1$ (KP-I) can be obtained  when surface tension effects are large (see, for example, \cite{Ablowitz2} and references therein). Useful transformations  mapping solutions of the cKP and KP equations were independently found in \cite{Johnson1,Johnson2} and \cite{LMS}.  They have been used to construct some special solutions of the cKP equation in \cite{Step1} and  \cite{KMS}. Indeed, the map
$$
W(\tau, \chi, V) \to U(\tau, \xi, Y) := W\left (\tau, \xi + \frac{Y^2}{12 \sigma^2 \tau}, \frac{Y}{\tau}\right )
$$
transforms any solution of the cKP equation (\ref{cKP}) into a solution of the KP equation  (\ref{KP}). Conversely, the map
$$
U(\tau, \xi, Y) \to W(\tau, \chi, V) := U\left (\tau, \chi - \frac{\tau V^2}{12 \sigma^2}, \tau V \right )
$$
transforms any solution of the KP equation (\ref{KP}) into a solution of the cKP equation (\ref{cKP}).   In \cite{Step2} it was pointed out that the transformations map rather general classes of evolution equations, containing the KP and cKP equations (as well as mapping their one-dimensional counterparts generalising the KdV and cKdV equations into some classes of solutions of the two-dimensional equations).

Another interesting transformation linking the cKP and KP equations was found in \cite{Lug} (see also \cite{Step1}). However, this transformation maps bounded solutions of the KP equation into unbounded solutions of the cKP equation. We do not consider transformations of this type in our paper.

In this paper we derive a generalisation of the cKP equation (\ref{cKP}) for surface gravity waves, which can be written in the form
\begin{eqnarray}
&&\left(H_\tau + 6 H H_\zeta +  H_{\zeta \zeta \zeta} +  \frac{\tau}{2 (\tau^2 - a^2)} H \right . \nonumber \\
&& \left . - \frac{\nu^2 a^2}{12 \sigma^2 (\tau^2 - a^2)}  H_\zeta   \right) _\zeta +  \frac{3 \sigma^2}{\tau^2-a^2} H_{\nu \nu}= 0, \label{ECKP}
\end{eqnarray}
where $\sigma^2 =\pm1$, describing waves with nearly-elliptic front.  This {\it elliptic cylindrical KP equation (ecKP)}  is derived from the full set of Euler equations for an incompressible fluid and free surface and rigid bottom boundary conditions (\ref{Euler}), written in the elliptic cylindrical coordinate system.   The linear long-wave equation, written in these coordinates, does not allow for {\it exact} solutions describing waves with elliptic front. However, there exists an {\it asymptotic} reduction to the necessary equation, and we show that this allows one to derive a generalisation of the cKP equation.

We chose to derive the ecKP equation from the Euler equations rather 
than using the velocity potential formulation, since this opens the 
way to the study of  internal and surface waves on a current for a fluid with arbitrary stratification, as well as accounting for the effects of a variable background and Earth's rotation (see \cite{GOSS, Grimshaw2} and references therein, for studies in the Cartesian geometry), which constitute rotational flows.

 We find transformations between KP, cKP and ecKP equations,
 generalising the transformations between KP and cKP equations in \cite{Johnson1,Johnson2,LMS}, and use them to construct some important special classes of solutions of the derived version of the Kadomtsev-Petviashvili equation for both cases (i.e. for ecKP-I and ecKP-II). Indeed, the map
\begin{eqnarray*}
&&U(\tau, \xi, Y) \to  \nonumber \\
&&H(\tau, \zeta, \nu) := U \left  (\tau, \zeta - \frac{\tau \nu^2}{12 \sigma^2}, \sqrt{\tau^2 - a^2} \nu \right )
\end{eqnarray*}
transforms any solution of the KP equation into a solution of the ecKP equation. Conversely, the map 
\begin{eqnarray*}
&&H(\tau, \zeta, \nu) \to \\
&&U(\tau, \xi, Y) := H \left  (\tau, \xi + \frac{1}{12 \sigma^2} \frac{\tau Y^2}{\tau^2 - a^2}, \frac{Y}{\sqrt{\tau^2 - a^2}} \right )
\end{eqnarray*}
transforms any solution of the ecKP equation into a solution of the KP equation.

 The ecKP equation (\ref{ECKP}) derived in our paper is an integrable model, which can be obtained as a compatibility condition of the following linear problem (Lax pair):
{\small \begin{eqnarray}
&&\sigma\psi_\nu=\sqrt{\tau^2-a^2}\psi_{\zeta\zeta} +\left(\sqrt{\tau^2-a^2}H(\tau,\zeta,\nu) \right.  \nonumber \\
&&\left.  -\dfrac{\tau\zeta}{12\sqrt{\tau^2-a^2}}+ \dfrac{a^2\nu^2}{144\sigma^2\sqrt{\tau^2-a^2}}\right)\psi, \label{la1}\\
&&\psi_\tau=-4\psi_{\zeta\zeta\zeta}-\left(6H(\tau,\zeta,\nu)- \dfrac{a^2\nu^2}{12\sigma^2(\tau^2-a^2)}\right)\psi_\zeta- \nonumber \\
&& -\left(3H_\zeta(\tau,\zeta,\nu)+\dfrac{3\sigma \wtH(\tau,\zeta,\nu)}{\sqrt{\tau^2-a^2}} -\dfrac{a^2\zeta\nu}{12\sigma(\tau^2-a^2)^{3/2}}\right)\psi. \qquad \label{la2}
\end{eqnarray}}
Indeed, the compatibility conditions have the form
\begin{eqnarray*}
&& \wtH_\zeta=H_\nu, \\
&& H_\tau+6HH_\zeta+H_{\zeta\zeta\zeta} +\dfrac{\tau}{2(\tau^2-a^2)}H \\
&& -\dfrac{a^2\nu^2}{12\sigma^2(\tau^2-a^2)}H_\zeta +\dfrac{3\sigma^2}{\tau^2-a^2}\wtH_\nu=0.
\end{eqnarray*}
When  $a=0$ we recover the Lax pair of the cKP equation \cite{Dryuma} (see also \cite{KMS}).

A solution $\psi_{ecKP}(\tau,\zeta,\nu)$ of the linear system  \eqref{la1}, \eqref{la2} is expressed via the solution  $\psi_{KP}(\tau,\xi,Y)$ of the linear system of the KP equation ($\wtU_\xi= U_Y$)
\begin{align*}
&\sigma\psi_{Y}=\psi_{\xi\xi}+U(\tau,\xi,Y)\psi,\\
&\psi_\tau=-4\psi_{\xi\xi\xi}-6U(\tau,\xi,Y)\psi_\xi \\
&-(3U_\xi(\tau,\xi,Y) +3\sigma\wtU(\tau,\xi,Y))\psi
\end{align*}
as follows
{\small \begin{eqnarray*}
&&\psi_{ecKP}(\tau,\zeta,\nu)=\psi_{KP}\left(\tau,\zeta-\dfrac{\tau\nu^2}{12\sigma^2}, \sqrt{\tau^2-a^2}\nu\right) \\
&&\times \exp\left\{-\dfrac{\zeta\tau\nu}{12\sigma\sqrt{\tau^2-a^2}} +\dfrac{\nu^3(\tau^2+a^2)}{432\sigma^3\sqrt{\tau^2-a^2}}\right\}.
\end{eqnarray*}}
Let us also note that the functions $\wtH$ and  $\wtU$ are related by
\begin{equation*}
\wtH(\tau,\zeta,\nu)=\sqrt{\tau^2-a^2}\,\wtU\left(\tau,\xi, Y\right)-\dfrac{\tau\nu}{6\sigma^2}U\left(\tau,\xi, Y\right).
\end{equation*}
The Jacobian of the transformation 
$$\xi=\zeta-\tau\nu^2/(12\sigma^2), \quad Y=\sqrt{\tau^2-a^2}\nu,
$$
 used to construct some special solutions of the ecKP in the following sections, is equal to $\sqrt{\tau^2-a^2}$. It is positive and bounded for all $\tau>a$. (This condition is automatically satisfied for our derivation.)
 
  The ecKP equation is an integrable equation containing an arbitrary parameter $a$, and it reduces to the cKP equation both when this parameter tends to zero, and when $\tau \gg a$. The latter is the mathematical manifestation of the intuitively clear physical phenomenon: wave fronts will become nearly-concentric far away from the wave sources or boundaries, while they will ``remember" their geometrical shape during the intermediate evolution.

 The KP and cKP equations belong to the family of {\it universal} integrable models of modern nonlinear physics (see, for example, \cite{AS, ZMNP}). As well as for surface waves, these equations have been derived in many physical settings (see for example references in \cite{Ablowitz2}), including internal waves in a stratified fluid \cite{Grimshaw, Lipovskii} and, most recently, matter waves in Bose-Einstein condensates (BEC) (see \cite{Huang, TDP} and references therein) and cosmic dust-ion-acoustic waves \cite{GT1}.
 Thus, the new version of the equation derived in this paper could find many useful applications to the description of the wave motion in problems where sources, boundaries and obstacles have elliptic or nearly-elliptic  geometry.

Our paper is organised as follows. In Section II (with Appendix A) we describe the derivation of the ecKP equation in the  context of the classical surface gravity waves problem for an incompressible fluid. From mathematical perspective, the derivation for water wave problems is more challenging than similar derivations for problems where nonlinear and dispersive terms are present in the equations, rather than originating from the free surface boundary conditions.The equation can be readily derived in other physical contexts.  In Section III we find transformations between arbitrary solutions of the derived version of the KP equation and the original KP equation. Section IV (with Appendix B) is devoted to the lumps, line solitons and quasiperiodic solutions of the ecKP-I and ecKP-II equations.  In Section V we discuss the approximate solutions for surface waves described by the derived equation. We conclude in Section VI by outlining possible applications and generalisations of our results.


\section{Derivation of the elliptic cylindrical KP equation}

We consider the classical water wave problem for an incompressible fluid, described by the full set of Euler equations with free surface and rigid horizontal bottom boundary conditions (\ref{Euler}).
Since we aim to consider waves with the nearly-elliptic front, we write this set of equations in the elliptic cylindrical coordinate system:
\begin{eqnarray*}
&&x = d \cosh \alpha \cos \beta, \\
&&y = d \sinh \alpha \sin \beta, \\
&&z = z,
\end{eqnarray*}
where the dimensional parameter $d$ has the meaning of  half of the distance between the foci of the coordinate lines, and change the two horizontal components of the velocity vector appropriately:
\begin{eqnarray*}
u \to  u \cos \beta - v \sin \beta, \quad
 v \to u \sin \beta + v \cos \beta.
\end{eqnarray*}
Here, we keep the same notations $(u, v, w)$ for the projections of the velocity vector on the new coordinate lines.

Let $z = h_0$ be the unperturbed fluid depth, $\lambda$ be the characteristic wavelength, $p_a$ the atmospheric pressure, and $h_s$ the characteristic free surface elevation. We nondimensionalise the variables
\begin{eqnarray*}
&&x \to \lambda x, \quad y \to \lambda y, \quad z \to h_0 z, \quad t \to \frac{\lambda}{\sqrt{gh_0}} t, \\
&&u \to \sqrt{gh_0} u, \quad v \to \sqrt{g h_0} v, \quad w \to \frac{h_0 \sqrt{gh_0}}{\lambda} w, \\[2ex]
 &&h \to h_0 + h_s \eta, \quad p \to p_a + \rho g (h_0-z) + \rho g h_0 p,
\end{eqnarray*}
which leads to the appearance of two usual nondimensional parameters in the problem: the long wavelength parameter
$
\delta = \frac{h_0}{\lambda},
$
and the small amplitude parameter
$
\epsilon = \frac{h_s}{h_0},
$
as well as a new nondimensional parameter
$
\gamma = \frac{d}{\lambda},
$
which is not necessarily small.
Scaling the dependent variables
$$
u \to \epsilon u, \quad v \to \epsilon v, \quad w \to \epsilon w, \quad p \to \epsilon p,
$$
we bring the full set of Euler equations in the elliptic cylindrical coordinates to the form:
{\small
\begin{eqnarray}
&&u_t + \epsilon \left [ w u_z  + \frac{E (uu_{\alpha} + v u_{\beta} - v^2) + F (v u_{\alpha} - u u_{\beta} + uv)}{2 \gamma e^{\alpha} G} \right ]  \nonumber \\
&& = - \frac{E p_{\alpha} - F p_{\beta}}{2 \gamma e^{\alpha} G}, \label{1} \\
&&v_t + \epsilon \left [ w v_z + \frac{E (uv_{\alpha} + v v_{\beta} + uv) + F (v v_{\alpha} - u v_{\beta} - u^2)}{2 \gamma e^{\alpha} G} \right ]  \nonumber \\
&&= - \frac{ E p_{\beta} + F p_{\alpha}}{2 \gamma e^{\alpha} G}, \label{2} \\[2ex]
&&\delta^2 \left [w_t + \epsilon \left \{ w w_z + \frac{E (uw_{\alpha} + v w_{\beta}) + F (v w_{\alpha} - u w_{\beta})}{2 \gamma e^{\alpha} G} \right \} \right]  \nonumber \\
&& = -p_z, \label{3} \\[2ex]
&&w_z + \frac{E (u_{\alpha} + v_{\beta} + u) + F (v_{\alpha} - u_{\beta} + v)}{2 \gamma e^{\alpha} G}  = 0, \label{4} 
\end{eqnarray}
}
{\small
\begin{eqnarray}
&&p|_{z = 1 + \epsilon \eta (\alpha, \beta, t) } = \eta  -  W_e \   \delta^2\  \left \{ \eta_{\alpha \alpha} + \eta_{\beta \beta} \right. \nonumber \\
&& \left. + \epsilon^2 \frac{\delta^2}{\gamma^2} \left [ \frac{\eta_\beta^2 \eta_{\alpha \alpha} + \eta_\alpha^2 \eta_{\beta \beta} - 2 \eta_{\alpha} \eta_{\beta} \eta_{\alpha \beta} }{G} \right. \right. \nonumber  \\
&&  \left. \left. +  \frac{\eta_{\alpha}  \sinh 2 \alpha + \eta_{\beta} \sin 2 \beta }{2 G^2}\right ] \right \} / \nonumber \\
&&  \left \{ \gamma^2 G \left ( 1 +  \epsilon^2 \frac{\delta^2}{\gamma^2} \frac{\eta_{\alpha}^2 + \eta_{\beta}^2}{G} \right )^{3/2} \right \} ,  \label{5} \\[2ex]
&&w|_{z = 1 + \epsilon \eta (\alpha, \beta, t) } = \eta_t   + \epsilon  \frac{E (u \eta_{\alpha} + v \eta_{\beta}) + F (v \eta_{\alpha} - u \eta_{\beta})}{G}, \qquad
\label{6}   \\[2ex]
&&w|_{z=0} = 0.  \label{7}
\end{eqnarray}
}
Here, $W_e = \frac{\Gamma}{\rho g h_0^2}$ is the Weber number, and we denoted
\begin{eqnarray*}
&&E = e^{2\alpha} - \cos 2 \beta,  \quad F = \sin 2\beta, \\
&&G = \sinh^2 \alpha + \sin^2 \beta.
\end{eqnarray*}

This set of equations reduces to the Euler equations written in cylindrical coordinates in the limit
\begin{equation}
\alpha \to \infty, \quad \gamma \to 0 \quad \mbox{with} \quad \frac 12 \gamma e^{\alpha} \to r \quad \mbox{being finite}.
\label{8}
\end{equation}

The equation for linear waves (in the long-wave approximation) is  easily obtained from equations (\ref{1}) - (\ref{7}) with $\epsilon = \delta = 0$ as
\begin{equation}
\eta_{tt} = \frac{\eta_{\alpha \alpha} + \eta_{\beta \beta}}{\gamma^2 (\sinh^2 \alpha + \sin^2 \beta)}.
\label{9}
\end{equation}
Note that equation (\ref{9}) indeed reduces to the equation
\begin{equation}
\eta_{tt} - (\eta_{rr} + \frac 1r \eta_r + \frac{1}{r^2} \eta_{\beta \beta}) = 0
\label{10}
\end{equation}
for the long linear waves in the polar cylindrical coordinates in the limit (\ref{8}).  The derivation of the cylindrical KP (cKP) equation  (also known as the nearly-concentric KP equation \cite{Johnson1, Johnson2}) is based on the existence of solutions of (\ref{10}), which do not depend on $\beta$, i.e. there exists an exact reduction of the equation (\ref{10}) to the equation
$$
\eta_{tt} - (\eta_{rr} + \frac 1r \eta_r) = 0.
$$
Unlike (\ref{10}), equation (\ref{9}) does not have an {\it exact} reduction to the equation with no dependence on $\beta$, which would seem necessary in order to derive a version of the KP equation for waves with nearly-elliptic front. Nevertheless, such an equation exists as an {\it asymptotic} reduction, and it turns out that this allows for a generalisation of the cKP equation to be derived.

Next, we introduce the variables
\begin{eqnarray*}
&&\zeta = \frac{\epsilon^2}{\delta^2} \left ( \gamma \cosh \alpha - t \right ),  \\
&&R = \frac{\epsilon^6}{\delta^4} \gamma \cosh \alpha, \quad \nu =  \frac{\delta}{\epsilon^2} \sin \beta, \\
&& u =  \frac{\epsilon^3}{\delta^2} U, \quad   v =  \frac{\epsilon^5}{\delta^3}  V, \quad w = \frac{\epsilon^5}{\delta^4}  W, \\
&&\eta = \frac{\epsilon^3}{\delta^2} H, \quad p = \frac{\epsilon^3}{\delta^2} P.
\end{eqnarray*}
which generalise a change of variables for the cylindrical coordinates \cite{Johnson1}.
We use a large distance variable $R$ in preference to large time, but one can also work throughout using an analogous large time variable, $T = \frac{\epsilon^6}{\delta^4} t$.
Here, $2 \gamma \cosh \alpha$ is the nondimensional sum of the distances from a point on an ellipse to its foci. Thus, $\zeta$ is an {\it asymptotic} characteristic coordinate for waves with nearly-elliptic front, and it becomes the characteristic coordinate for the concentric waves in the limit (\ref{8}). Note that in this derivation the variable $\nu$ is proportional to $\sin \beta$ and not just $\beta$, unlike the derivation for the concentric waves \cite{Johnson1, Johnson2}. This increases the range of the formal asymptotic validity of the model.

In these variables, the problem formulation (\ref{1}) - (\ref{7}) assumes the form containing a single small parameter
$
\Delta = \frac{\epsilon^4}{\delta^2}, 
$
and a non-dimensional parameter
$
A = \gamma \frac{\epsilon^6}{\delta^4},
$
which is not necessarily small. The equations are given in Appendix A.

We now seek an asymptotic solution of this system of equations and boundary conditions  in the form
$$
H = H_0 + \Delta H_1 + O(\Delta^2),
$$
with similar expansions for $U, V, W$ and $P$. At leading order ($O(1)$) we obtain
\begin{eqnarray*}
&&U_{0\zeta} = P_{0\zeta}, \\
&&V_{0\zeta} = \frac{1}{\sqrt{R^2-A^2}} P_{0\nu} + \frac{R-\sqrt{R^2-A^2}}{\sqrt{R^2-A^2}} \nu P_{0\zeta}, \\
&&P_{0z} = 0, \quad U_{0\zeta} + W_{0z} = 0, \\
&&P_0|_{z=1} = H_0, \quad W_0|_{z=1} = -H_{0\zeta}, \quad W_0|_{z=0} = 0,
\end{eqnarray*}
which yields, imposing the condition that the perturbation in $U$ is caused only by the passing wave,
\begin{eqnarray}
&&P_0 = H_0, \quad U_0 = H_0, \quad W_0 = - H_{0\zeta} z, \label{21} \\
&&V_{0\zeta} =  \frac{1}{\sqrt{R^2-A^2}} H_{0\nu} + \frac{R-\sqrt{R^2-A^2}}{\sqrt{R^2-A^2}} \nu H_{0\zeta}.
\label{22}
\end{eqnarray}

At the next order ($O(\Delta)$) we obtain the following equations and boundary conditions:
{\small
\begin{eqnarray}
&&U_{1\zeta} - P_{1\zeta} = P_{0R} + U_0 U_{0\zeta}  + W_0 U_{0z}  \nonumber \\
&&- \frac{R - \sqrt{R^2-A^2} }{R^2 - A^2} \left (\nu P_{0\nu} + R  \nu^2 P_{0\zeta} \right ), \label{23} \\[2ex]
&&V_{1\zeta} - \frac{1}{\sqrt{R^2-A^2}} P_{1\nu}  - \frac{R - \sqrt{R^2-A^2} }{\sqrt{R^2 - A^2}} \nu  P_{1\zeta} \nonumber \\
&&= U_0 V_{0\zeta} + W_0 V_{0z}  +  \frac{R - \sqrt{R^2-A^2} }{\sqrt{R^2 - A^2}} \nu  P_{0R} \nonumber \\
&&+ \frac{(R - \sqrt{R^2 - A^2}) (R^2+A^2)}{2 (R^2 - A^2)^{3/2}} \nu^3 P_{0\zeta}  \nonumber  \\
&& -\frac{(R-\sqrt{R^2-A^2})^2 + R^2}{2 (R^2-A^2)^{3/2}} \nu^2 P_{0\nu}, \label{24} \\[2ex]
&&P_{1z} =  W_{0\zeta}, \label{25} \\[2ex]
&&U_{1\zeta} + W_{1z} = - U_{0R} + \frac{R - \sqrt{R^2-A^2}}{R^2-A^2} (R \nu^2 U_{0\zeta} + \nu U_{0\nu})  \nonumber \\
&&  - \frac{1}{\sqrt{R^2-A^2}} \left ((R-\sqrt{R^2-A^2}) \nu V_{0\zeta} +  V_{0\nu} + U_0\right ), \label{26} \\[2ex]
&&P_1|_{z=1} + H_0 P_{0z}|_{z=1} = H_1 - W_e\  H_{0\zeta \zeta}, \label{27}\\[2ex]
&&W_1|_{z=1} + H_0 W_{0z}|_{z=1} = -H_{1\zeta} + U_0 H_{0\zeta}, \label{28}\\[2ex]
&&W_1|_{z=0} = 0.\label{29}
\end{eqnarray}
}
Then, (\ref{21}), (\ref{25}) and (\ref{27}) yield
$$
P_1 = -H_{0\zeta \zeta} \left ( \frac{z^2-1}{2} + W_e \right ) + H_1,
$$
hence from (\ref{23}), (\ref{26}) and (\ref{29}) we find, using (\ref{21}), that
{\small
\begin{eqnarray}
&&W_1 = H_{0\zeta \zeta \zeta} \left [ \frac{z^3}{6} + \left (W_e - \frac{1}{2} \right ) z \right ] \nonumber \\
&& -  \left [H_{1\zeta} + 2 H_{0R} + H_0 H_{0\zeta}
- \frac{R - \sqrt{R^2-A^2}}{R^2-A^2} \left ( 2 \nu H_{0\nu} \right . \right . \nonumber  \\
&& + \left . \left . 2R \nu^2 H_{0\zeta}  \right ) + \frac{1}{\sqrt{R^2-A^2}} \left ((R-\sqrt{R^2-A^2}) \nu V_{0\zeta} \right . \right. \nonumber \\
&&\left.  \left . + V_{0\nu} + H_0\right ) \right ] z.
\end{eqnarray}
}
Finally, substituting $W_1$ into the remaining boundary condition (\ref{28}),  differentiating with respect to $\zeta$ and using (\ref{22}), we obtain the {\it elliptic cylindrical KP (ecKP) equation} 
\begin{eqnarray}
&&\left [2 H_{0R} + 3 H_0 H_{0\zeta} +\left ( \frac 13 - W_e \right ) H_{0\zeta \zeta \zeta}+  \frac{R}{R^2-A^2} H_0  \right. \nonumber \\
&&\left. - A^2 \frac{\nu^2}{R^2-A^2} H_{0\zeta}  \right ]_{\zeta} + \frac{1}{R^2-A^2} H_{0\nu \nu} = 0.
\label{ecKP}
\end{eqnarray}
Note that equation (\ref{24}), written for the completeness of the set of equations, allows one to find $V_1$ and is not used in the derivation of the ecKP equation.

The scaling transformation
$$
\tilde \alpha R \to \tau, \quad \tilde \beta \zeta \to \zeta, \quad \tilde \gamma \nu \to \nu, \quad H_0 \to \tilde \delta H
$$
where $\tilde \alpha$ is a free parameter and
\begin{eqnarray*}
&&\tilde \beta = \left (\frac{2 \tilde \alpha}{\frac 13 - W_e} \right )^{1/3}, \\
&&\tilde \gamma = (6 \tilde \alpha \tilde \beta \sigma^2)^{1/2}, \\
&& \tilde \delta = 4 \frac{\tilde \alpha} {\tilde \beta}, \\
&& \sigma^2 = {\rm sign}\  (\tilde \alpha \tilde \beta)
\end{eqnarray*}
brings the derived equation (\ref{ecKP}) to the form
\begin{eqnarray*}
&&\left(H_\tau + 6 H H_\zeta +  H_{\zeta \zeta \zeta} +  \frac{\tau}{2 (\tau^2 - a^2)} H \right. \\
&&\left. - \frac{a^2 \nu^2}{12 \tilde \alpha^2 \sigma^2 (\tau^2 - a^2)}  H_\zeta   \right) _\zeta +  \frac{3 \sigma^2}{\tau^2-a^2} H_{\nu \nu}= 0,
\end{eqnarray*}
shown in the Introduction. Here, $a = \tilde \alpha A$. If we let $\tilde \alpha = 1$, then $a = A$ and $\sigma^2 = {\rm sign}\ \left (\frac 13 - W_e \right)$. For typical water waves, $\sigma^2 = 1$ ($W_e < \frac 13$). However, $\sigma^2 = -1$ if the effects of surface tension are strong ($W_e > \frac 13$). It is natural to call the corresponding equations {\it ecKP-II} and {\it ecKP-I}, respectively, similarly to the terminology used in the Cartesian geometry.

\section{Transformations between KP, cKP and ecKP equations}

Considerations used to find the mapping from the solutions of the KdV equation to the class of solutions of the cKP equation \cite{Johnson1, Johnson2} can be extended to obtain transformations between arbitrary solutions of all three versions of the KP equation, related to the Cartesian, cylindrical and elliptic cylindrical coordinates, respectively. The resulting transformations generalise the transformations between the KP and cKP equations \cite{Johnson1,Johnson2,LMS}, discussed in the Introduction.

Indeed, the geometry of a wave with nearly-elliptic front, considered simultaneously in the Cartesian and  elliptic cylindrical coordinates, suggests the introduction of the sum and the difference of the nondimensional distances from a point on the wave front to the two foci of the coordinate system
\begin{eqnarray*}
&&d_1 + d_2 = 2 \gamma \cosh \alpha, \\
&&d_1 - d_2 = 2 \gamma \cos \beta,
\end{eqnarray*}
where the foci have the following Cartesian coordinates:
$
F_1 (-\gamma, 0)
$
and
$
F_2 (\gamma, 0).
$
We recall that the variables have been nondimensionalised, as discussed in section 2, and $\gamma = \frac{d}{\lambda}$. Note that $\frac 12 (d_1 + d_2) - t$ corresponds, up to the scaling, to the asymptotic characteristic variable $\zeta$, introduced in section 2.

Then, for  the area satisfying $\frac{y}{x-\gamma}, \frac{y}{x+\gamma} \to 0$, we obtain the following asymptotic behaviour
\begin{eqnarray*}
&&\frac 12 (d_1 + d_2) - t \\
&&= \frac 12 \left (\sqrt{(x+\gamma)^2 + y^2} + \sqrt{(x-\gamma)^2 + y^2}\right ) - t \\
&&\sim x-t + \frac 14 y^2 \left (\frac{1}{x+\gamma} + \frac{1}{x-\gamma}\right ).
\end{eqnarray*}
Next, for sufficiently large $\alpha$ and small $\beta$, our nondimensional variable
$
x = \gamma \cosh \alpha \cos \beta \sim \frac{\delta^4}{\epsilon^6} R,
$
and the previous asymptotics can be rewritten as
$$
\frac 12 (d_1 + d_2) - t  \sim \xi+ \frac 12 Y^2 \frac{R}{R^2-A^2},
$$
where $\xi = x-t, Y = \frac{\epsilon^3}{\delta^2} y$ and $A = \gamma \frac{\epsilon^6}{\delta^4}$. Similarly,
$$
\frac 12 (d_1 - d_2)   \sim \gamma - \frac A2  \frac{Y^2}{R^2 - A^2}.
$$
This asymptotic behaviour of the geometrically meaningful objects  motivates the change of variables
$$
H_0 (R, \zeta, \nu) = \eta (R, \xi, Y),
$$
where
$$
\zeta = \xi + \frac 12 Y^2 \frac{R}{R^2 - A^2}, \quad \nu = \frac{Y}{\sqrt{R^2 - A^2}}.
$$
It is then verified by direct calculation that this transformation maps the ecKP equation (\ref{ecKP}) to the KP equation,
written in the form
$$
\left [2 \eta_{R} + 3 \eta \eta_{\xi} + \left ( \frac 13 - W_e \right ) \eta_{\xi \xi \xi} \right ]_{\xi} + \eta_{YY} = 0.
$$

To finish this section, let us summarise the transformations between all three versions of the KP equation. We write the KP equation in the canonical form
\begin{equation}
\left (U_{\tau} + 6 U U_{\xi} +  U_{\xi \xi \xi} \right )_{\xi} +3 \sigma^2  U_{YY} = 0,
\label{KP1}
\end{equation}
the cKP equation in the similar form
$$
\left (W_{\tau} +  6 W W_{\chi} +  W_{\chi \chi \chi} + \frac{1}{2 \tau} W  \right )_{\chi} + \frac{3 \sigma^2}{\tau^2} W_{VV} = 0,
$$
and the ecKP equation as
\begin{eqnarray}
&&\left (H_{\tau} + 6 H H_{\zeta} + H_{\zeta \zeta \zeta} +  \frac{\tau}{2 (\tau^2-a^2)} H \right. \nonumber \\
&&\left. -  \frac{a^2 \nu^2}{12 \sigma^2 (\tau^2-a^2)} H_{\zeta}  \right )_{\zeta} + \frac{3 \sigma^2}{\tau^2-a^2} H_{\nu \nu} = 0.
\label{ecKPa}
\end{eqnarray}
Then, the map
\begin{eqnarray*}
&&U(\tau, \xi, Y) \to \\
&&W(\tau, \chi, V) := U\left (\tau, \chi - \frac{\tau V^2}{12 \sigma^2},  \tau V \right )
\end{eqnarray*}
transforms any solution of the KP equation into a solution of the cKP equation,
and the map
\begin{eqnarray}
&&U(\tau, \xi, Y) \to  \nonumber \\
&&H(\tau, \zeta, \nu) := U \left  (\tau, \zeta - \frac{\tau \nu^2}{12 \sigma^2}, \sqrt{\tau^2 - a^2} \nu \right )
\label{eckpmap}
\end{eqnarray}
transforms any solution of the KP equation into a solution of the ecKP equation. Note, that the second transformation reduces to the first in the limit $a \to 0$. The map (\ref{eckpmap}) also shows that  for small $a$ and small values of $\tau$ any solution of the ecKP equation approaches some $Y$-independent solution of the KP equation.
These transformations can be inverted, and they can also be used to obtain the direct transformations between the cKP and ecKP equations.

 Indeed, the map inverting (\ref{eckpmap}) has the form
\begin{eqnarray*}
&&H(\tau, \zeta, \nu) \to \\
&&U(\tau, \xi, Y) := H \left  (\tau, \xi + \frac{1}{12 \sigma^2} \frac{\tau Y^2}{\tau^2 - a^2}, \frac{Y}{\sqrt{\tau^2 - a^2}} \right ).
\end{eqnarray*}
It transforms any solution of the ecKP equation into a solution of the KP equation. In particular, this map shows that for very large values of $\tau$ and finite values of $Y$ any solution of the ecKP equation will approach a $Y$-independent solution of the KP equation (possibly, a constant or zero). However, 
such large values of $\tau$ are likely to lie outside of the range of applicability of the derived model, and we do not discuss this limit any more.

The map
\begin{eqnarray*}
&&W(\tau, \chi, V) \to \\
&&H(\tau, \zeta, \nu) := W \left (\tau, \zeta - \frac{a^2 \nu^2}{12 \sigma^2 \tau}, \frac{\sqrt{\tau^2 - a^2}}{\tau} \nu \right )
\end{eqnarray*}
transforms any solution of the cKP equation into a solution of the ecKP equation, and the map
\begin{eqnarray*}
&&H(\tau, \zeta, \nu) \to \\
&&W(\tau, \chi, V) := H \left (\tau, \chi + \frac{a^2 \tau V^2}{12 \sigma^2 (\tau^2 - a^2)}, \frac{\tau}{\sqrt{\tau^2 - a^2}} V \right )
\end{eqnarray*}
transforms any solution of the ecKP equation into a solution of the cKP equation.

\section{Special solutions of ecKP-I and ecKP-II equations}

In this section we will consider some special solutions to the ecKP equation  (\ref{ecKPa}) {\it per se},
to illustrate the characteristic features of the equation. The considered
examples are exact solutions to the KP-I and KP-II equations: lumps, line solitons and
quasi-periodic solutions (see \cite{MS, BBEIM}), which become solutions to the ecKP-I and ecKP-II equations under the
map (\ref{eckpmap}).

If the $a$ in (\ref{eckpmap}) vanishes, the
ecKP solution reduces to the corresponding cKP solution. For small
``times" $\tau$, the cKP solutions look like solutions to the KdV equation
(essentially no dependence on the transversal variable), whereas they develop horseshoe-type
profiles for larger $\tau$. The ecKP solutions on the other hand show
such profiles already for small $\tau-a$ if $a>0$. For large $\tau$ ($\tau \gg a$) the
solutions tend asymptotically to the corresponding cKP solutions. We
will illustrate this behaviour at several examples.

\begin{figure}[h]
    \centering
    \includegraphics[width=8.5cm]{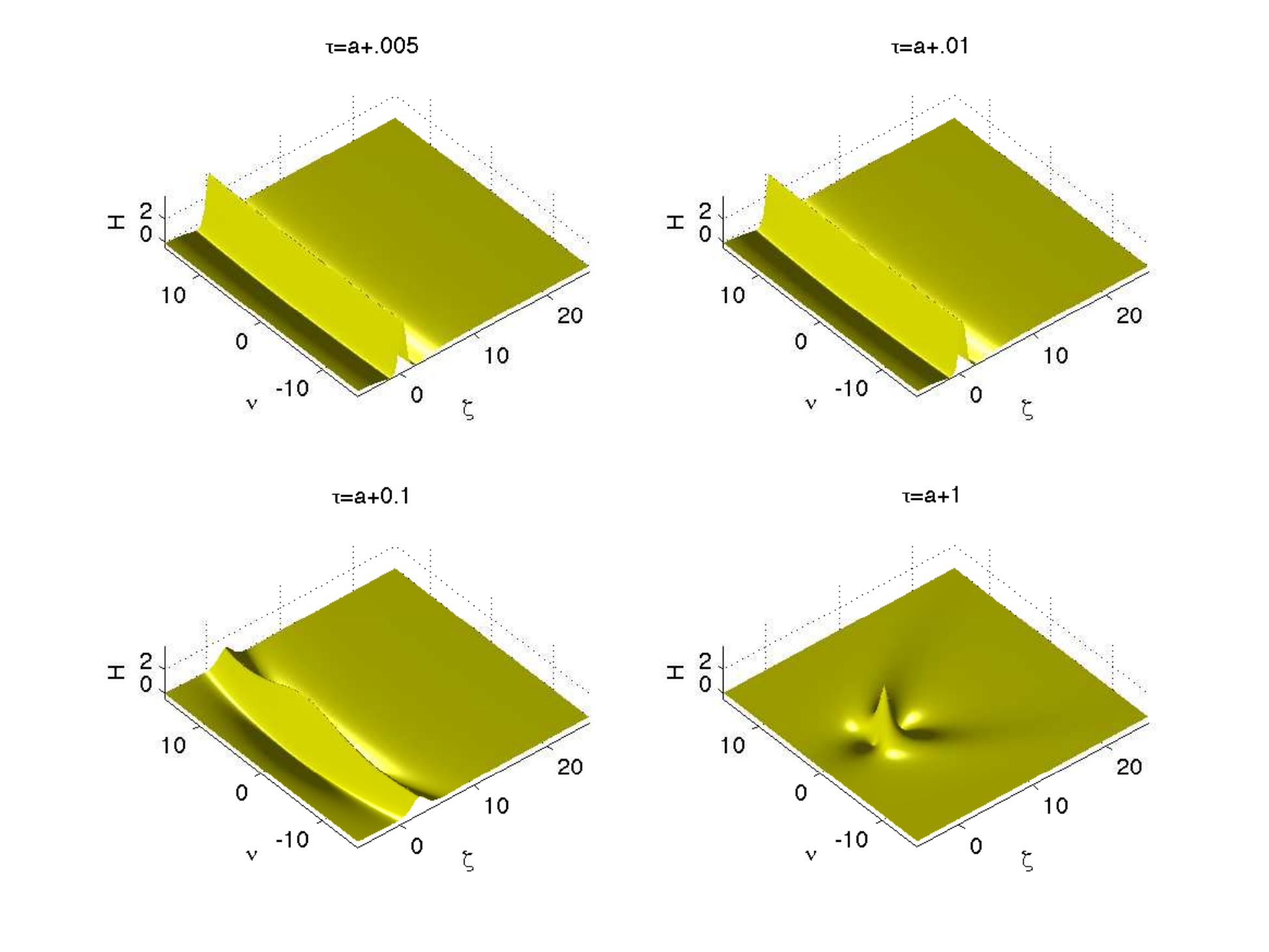}
\caption{Solution to the ecKP-I equation  obtained as the image of
the lump (\ref{lump}) with $\kappa = 1$ under the action of the map (\ref{eckpmap}) for $a=0.01$ and
several values of $\tau$.} \label{eckplumpa01}
\end{figure}

\begin{figure}[h]
    \centering
    \includegraphics[width=8.5cm]{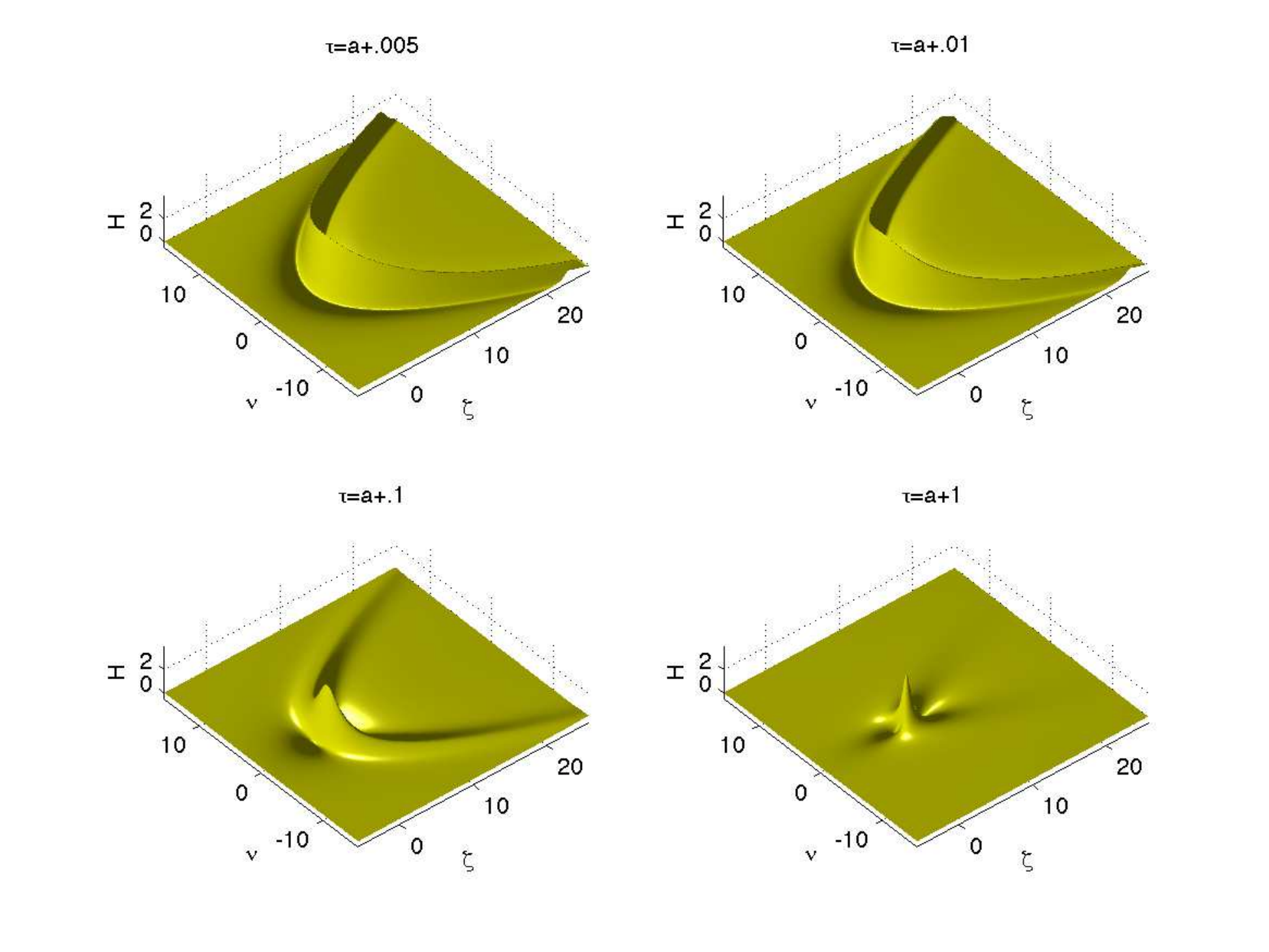}
\caption{Solution to the ecKP- I equation obtained as the image of
the lump (\ref{lump}) with $\kappa = 1$ under the action of the map (\ref{eckpmap}) for $a=1$ and
several values of $\tau$.} \label{eckplumpa1}
\end{figure}

The first example we consider is the KP-I  lump solution,
\begin{equation}
    U(\xi,Y,\tau) = \frac{4\kappa (1-\kappa (\xi-3\kappa  \tau)^{2}+\kappa ^{2}Y^{2})}{
    (1+\kappa (\xi-3\kappa \tau)^{2}+\kappa ^{2}Y^{2})^{2}},
    \label{lump}
\end{equation}
with $\kappa = 1$, under the map (\ref{eckpmap}). It is
visible that when $\tau \to a$ and $a$ is close to zero, the solution $H(\zeta, \nu, \tau)$ is essentially independent
of the coordinate $\nu$. This can be seen for $a=0.01$ in
Fig.~\ref{eckplumpa01}.
For larger values of $a$, the solution has a parabolic shape for
small $\tau - a$ as can be seen in
Fig.~\ref{eckplumpa1}.
\bigskip


%

\begin{figure}[h]
\centering
\includegraphics[width=8.5cm]{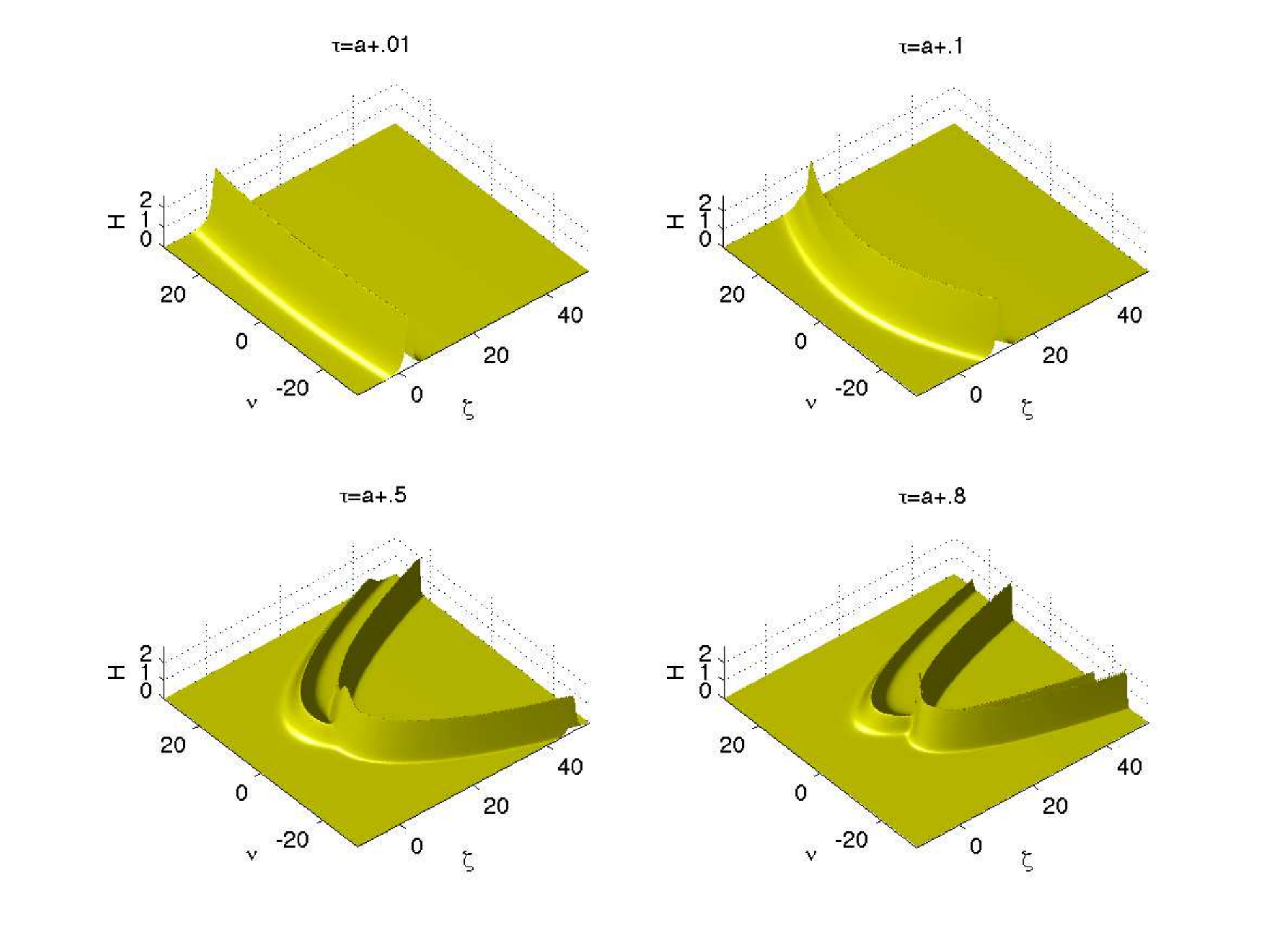}
\caption{2-soliton solution (\ref{2soliton})  of the ecKP-II equation
for $a=0.01$ with $k_{1}=1.5$, $k_{2}=0.5$, $k_{3}=-2$, $k_{4}=0$  for several values
of $\tau$.}\label{ckpsol2}
\end{figure}

\begin{figure}[h]
\centering
\includegraphics[width=8.5cm]{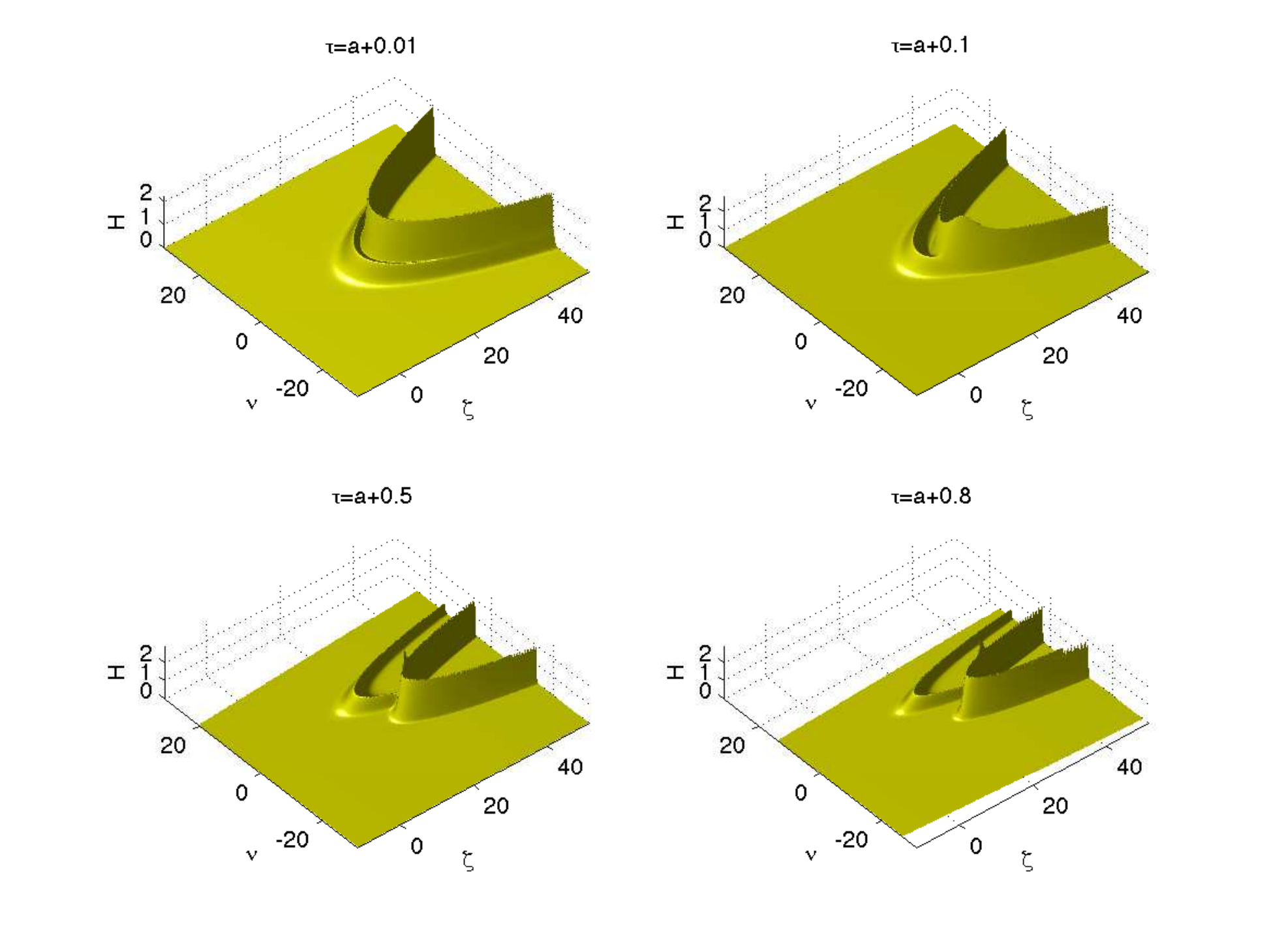}
\vspace{-0.5cm}
\caption{2-soliton solution (\ref{2soliton}) of the ecKP-II equation  for $a=1$ with
$k_{1}=1.5$, $k_{2}=0.5$, $k_{3}=-2$, $k_{4}=0$ for several values
of $\tau$.}\label{eckpsol2}
\end{figure}

Next, we consider the 2-soliton solution of the KP-II equation  in the form
\begin{equation}
U(\xi, Y, \tau)=2\partial_{x}^{2}\ln W\left(e^{\vartheta_{1}}+e^{\vartheta_{2}},
e^{\vartheta_{3}}+e^{\vartheta_{4}}\right),
\label{2soliton}
\end{equation}
where $\theta_j = k_j \xi + k_j^2 Y - 4 k_j^3 \tau$, $k_j$ are arbitrary constants, and $W$ is the Wronskian of the two functions.
It can be seen for $a=0.01$, i.e., close to the cKP case, in Fig.~\ref{ckpsol2}
where the formation of horseshoe waves can be clearly recognised.
The corresponding ecKP solution for $a=1$ is shown in
Fig.~\ref{eckpsol2} where the curved profiles are already present for
small $\tau - a$.


Quasi-periodic (multiphase) solutions of the ecKP equation can be obtained as the image of the known theta-functional solutions of the KP equation under the map (\ref{eckpmap}). The solutions are shown in Appendix B.

While the solutions of the ecKP equation are qualitatively similar to the solutions of the cKP equation, significant differences can be seen at the level of approximate solutions for the Euler equations, as shown in the next section. 


\section{Approximate solutions for surface waves}

Exact solutions of the derived equation allow us to obtain new asymptotic solutions for the classical water wave problem (\ref{Euler}).
In order to do that we return to the original nondimensional variables $x, y, t$ and re-parametrise our solution  as follows
{\small \begin{eqnarray*}
&&x = \gamma \cosh \alpha \cos \beta, \quad y = \gamma \sinh \alpha \sin \beta, \\
&& \eta = \frac{4}{6^{1/3}} (1-3 W_e)^{1/3} \sqrt{\frac{a}{\gamma}} H(\tau, \zeta, \nu),
\end{eqnarray*}}
where
\begin{eqnarray}
&&\left (H_{\tau} + 6 H H_{\zeta} + H_{\zeta \zeta \zeta} + \frac{\tau}{2 (\tau^2 - a^2)} H \right. \nonumber \\
&&\left.  - \frac{a^2 \nu^2}{12 \sigma^2 (\tau^2 - a^2)} H_{\zeta}\right  )_{\zeta} + \frac{3 \sigma^2}{\tau^2 - a^2} H_{\nu \nu} = 0,
\label{ecKPb}
\end{eqnarray}
 \begin{eqnarray*}
&&\sigma^2 = {\rm sign}\ (1 - 3 W_e)\quad \mbox{and}\quad  \tau = R = a \cosh \alpha,\\
&&\zeta = \frac{6^{1/3}  a}{\gamma \Delta (1 - 3 W_e)^{1/3}} (\gamma \cosh \alpha - t), \\
&&\nu = \frac{6^{2/3}}{ \Delta^{1/2} |1 - 3 W_e|^{1/6}} \sin \beta.
\end{eqnarray*}
Here, $t$ is the physical time (nondimensional). Below, we assume that  $t \ge 0$ and consider the initial stages of the evolution. We also have $\epsilon = \sqrt{\frac{\gamma}{a}} \Delta$ and $\delta = \frac{\gamma}{a} \Delta^{3/2}.$ Since asymptotic long-wave models usually provide a good qualitative (often quantitative) description even outside of the range of their formal asymptotic validity (i.e. the physical {\it applicability} of such models is usually wider than their {\it  formal asymptotic validity}), we plot the solutions for all $0 \le \beta < 2 \pi$ and $\alpha \ge 0$. Unless it is explicitly stated otherwise, at least parts of the shown solutions belong to the range of the formal validity of the asymptotic model (defined by $\tau \sim O(1), \zeta \sim O(1), \nu \sim O(1)$ as $\Delta \to 0$). 

The 1-soliton solution of the ecKP-II equation (i.e. the image of the 1-soliton solution of the KP-II equation under the map (\ref{eckpmap})) is explicitly written in the form
\begin{eqnarray}
&&H(\tau, \zeta, \nu) = \frac{K^2}{2}\  {\rm sech}^2 \left [ \frac{K}{2} \left (\zeta - \frac{\tau \nu^2}{12} + L \sqrt{\tau^2 - a^2} \nu \right. \right. \nonumber \\
&& \left. \left. - (K^2 + 3 L^2) \tau + \delta_0 \right ) \right ],
\label{H1}
\end{eqnarray}
where $K, L, \delta_0$ are arbitrary constants. In the examples shown below we let the Weber number $W_e = 0$, and  the phase shift $\delta_0 = 0$. It turns out that this single formula describes a variety of wave fronts. In what follows we provide the complete classification of these wave fronts, obtaining characteristic conditions on the parameters of the solutions (\ref{H1}) distinguishing various cases. We illustrate most of the wave fronts, plotting  the corresponding surface wave elevation $\eta$ for $\gamma = 1, a=2, \Delta = 1/2$. Similar solutions exist for the ecKP-I equation, as the image of an (unstable) line-soliton of the KP-I equation. We also note that an analogue of the solution (\ref{H1}) in cylindrical geometry (cKP) describes only a single type of a wave front (the picture is qualitatively similar to a part of the wave front shown in Fig.~\ref{sw1} below), and it can be plotted only for the limited values of the polar angle (even formally).

The wave obtained when $K=1, L=0$ is compact and symmetric, it is shown in Fig.~\ref{sw1} below.
\begin{figure}[htb]
    \centering
\includegraphics[width=8.5cm]{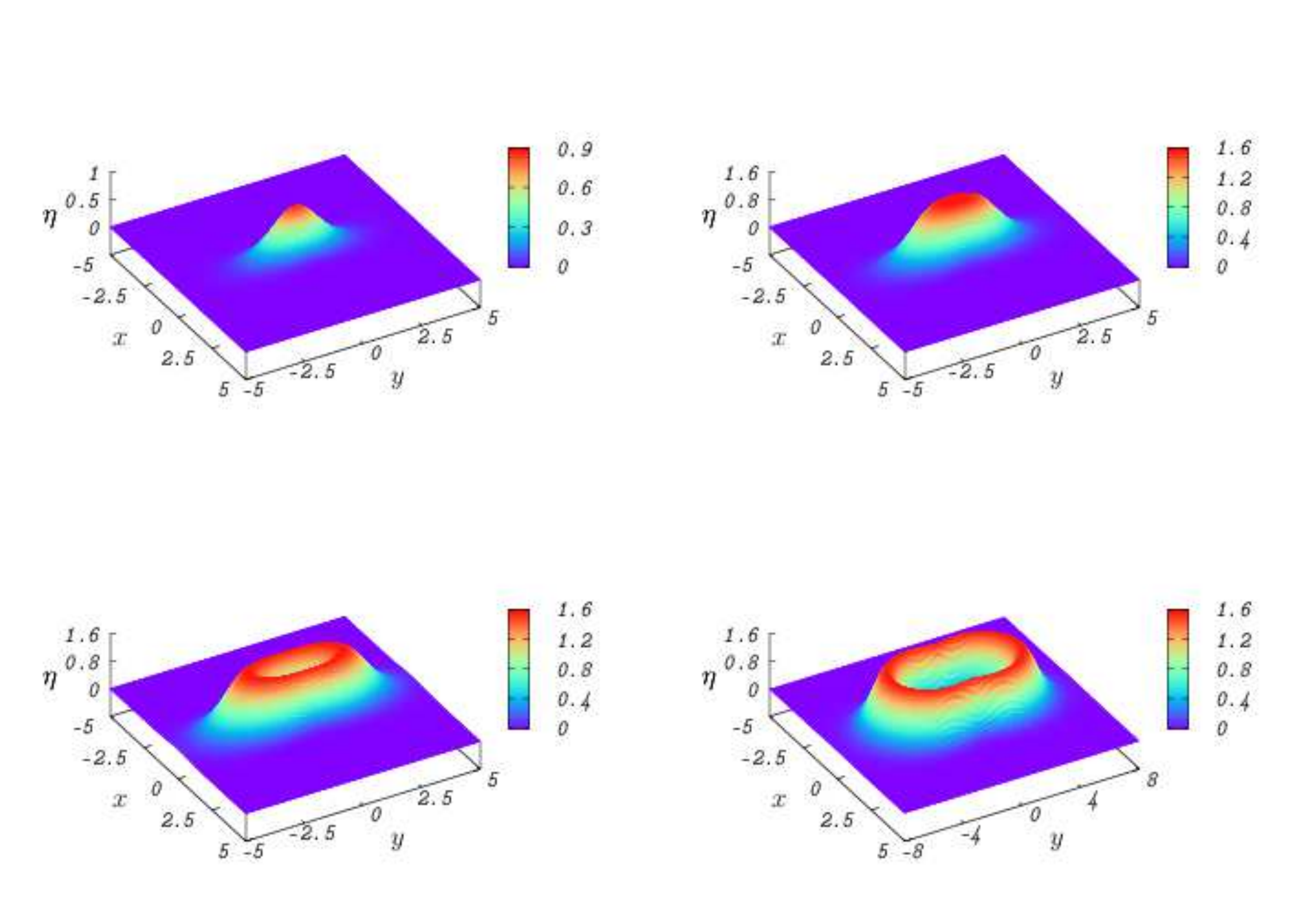}
\caption{Surface wave corresponding to the one-soliton solution  (\ref{H1}) of the ecKP-II equation with $K=1, L=0$ for $t=0$ (top left), $t=0.25$ (top right), $t = 0.5$ (bottom left), $t = 1$ (bottom right).}\label{sw1}
\end{figure}

\begin{figure}[htb]
    \centering
\includegraphics[width=8.5cm]{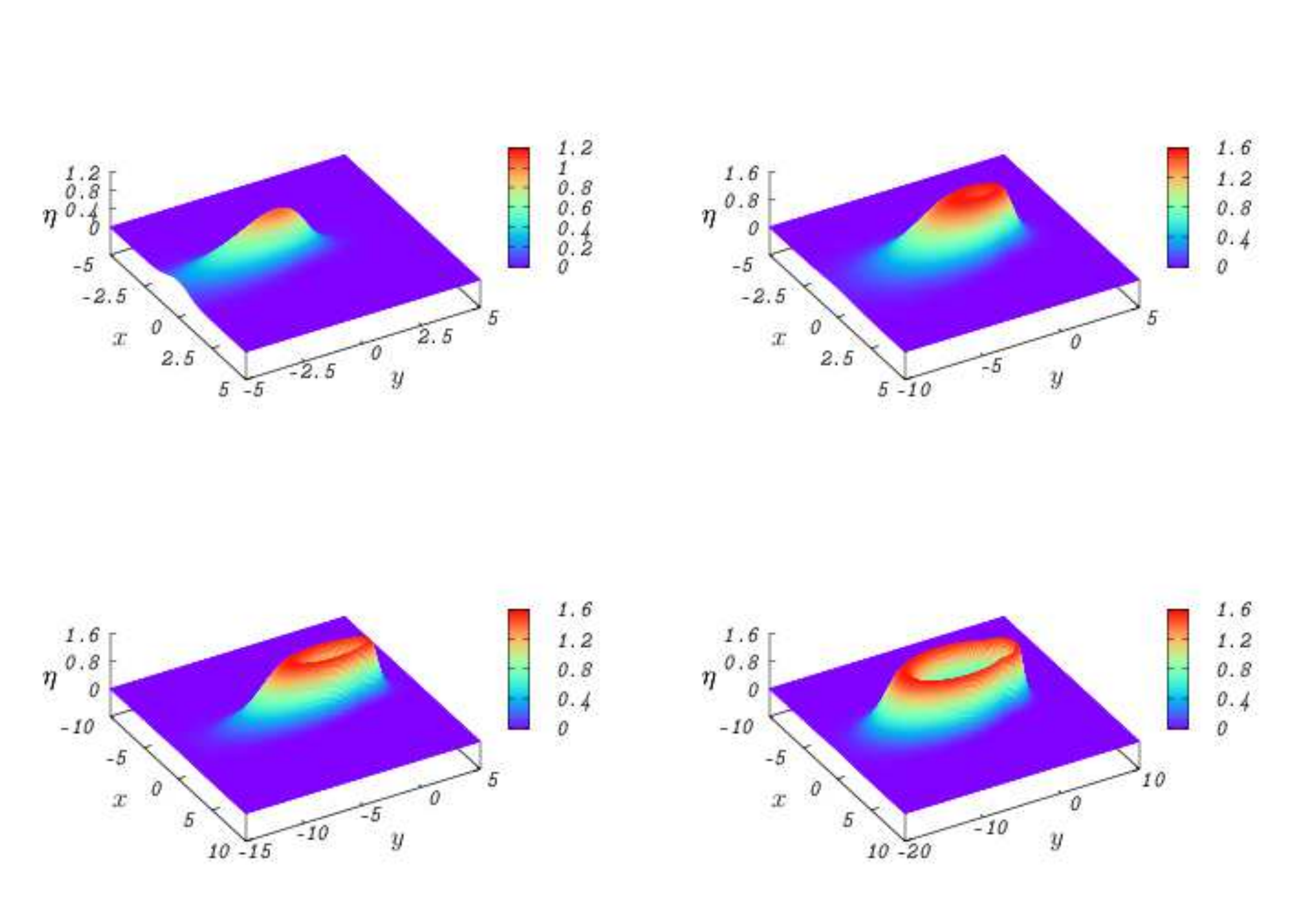}
\caption{Surface wave corresponding to the one-soliton solution (\ref{H1})  of the ecKP-II equation with $K=1, L=0.1$ for $t=0$ (top left), $t=0.25$ (top right), $t = 0.5$ (bottom left), $t = 1$ (bottom right).}\label{sw1.5}
\end{figure}

For $L\ne0$ the solution has no symmetry with respect to the $y$-coordinate, as one can see in Fig.~\ref{sw1.5} for $K=1$ and $L =0.1$. The change $L\to-L$ yields the reflection of the wave front with respect to the $x$-axis 
\begin{equation}
\eta(-L,x,y)=\eta(L,x,-y).
\label{L}
\end{equation}
 Therefore, it suffices to consider  $L>0$ or $L < 0$.

\begin{figure}[htb]
    \centering
\includegraphics[width=8.5cm]{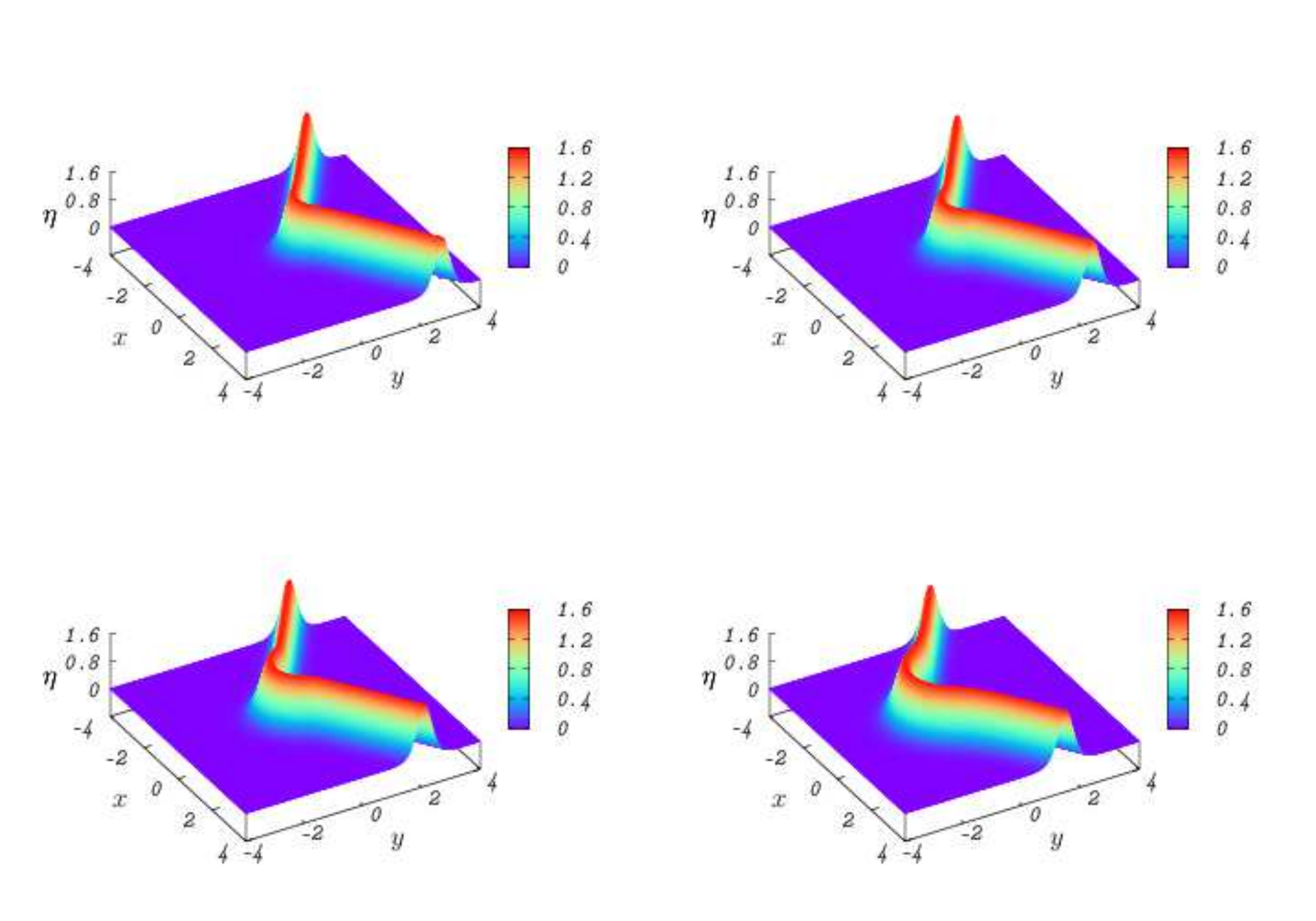}
\caption{Surface wave corresponding to the one-soliton solution  (\ref{H1})  of the ecKP-II equation  with $K=1, L=-0.5$ for $t=0$ (top left), $t=0.25$ (top right), $t = 0.5$ (bottom left), $t = 2$ (bottom right).}\label{sw2}
\end{figure}

When $|L|$ increases further,  the compact nearly - elliptic wave shape disappears. The wave becomes non-compact, and it rather describes the deformed line soliton, featuring an elliptic inhomogeneity in the central part of the wave. The solution is shown in Fig.~\ref{sw2} for $K=1, L=-0.5$.

For sufficiently large $|L|$ solution is localised in the vicinity of some point satisfying the relation $L\sin\beta=\abs{L}$, and strongly attenuates with time.
(The large values of $L$ lie outside of the range of validity of the model.)


In order to explain the observed features of the solution (\ref{H1}) and to obtain the corresponding conditions for the parameters of the solution, let us note that the maximum of its amplitude is attained when the argument of $\sech^2$ is equal to  zero (if this is possible):
\begin{equation*}
\zeta - \frac{\tau \nu^2}{12} + L \sqrt{\tau^2 - a^2} \nu - (K^2 + 3 L^2) \tau =0.
\end{equation*}
This condition can be written either as
\begin{eqnarray}
&&\left(2-\sin^2\beta-\dfrac{2\Delta}{6^{1/3}}\left[K^2+3L^2\right]\right)\cosh\alpha \nonumber \\
&&+2\Delta^{1/2}6^{1/3}L\sin\beta\sinh\alpha-\frac{2t}{\gamma}=0 \label{arg}
\end{eqnarray}
or as
\begin{eqnarray}
&&\sin^2\beta -2\Delta^{1/2}6^{1/3}L\tanh\alpha\sin\beta  \nonumber \\
&& +\frac{2\Delta}{6^{1/3}}\left[K^2+3L^2 \right] -2+ \frac{2t}{\gamma}\sech\alpha =0. \label{sin2b}
\end{eqnarray}

Let us first consider the case $L = 0$. For
\begin{equation*}
K^2\leq \frac{6^{1/3}}{2\Delta}
\end{equation*}
the solution has the form of a compact nearly-elliptic wave of narrowing width (as shown in Fig.~\ref{sw1}). Indeed,
\begin{equation*}
1-\dfrac{2\Delta}{6^{1/3}}K^2\leq \dfrac{2t}\gamma\sech \alpha\leq2-\dfrac{2\Delta}{6^{1/3}}K^2.
\end{equation*}
We note that although the width of the wave clearly changes, the amplitude is constant, which can be viewed as the manifestation of the {\it solitonic} nature of this solution.

\begin{figure}[htb]
    \centering
\includegraphics[width=8.5cm]{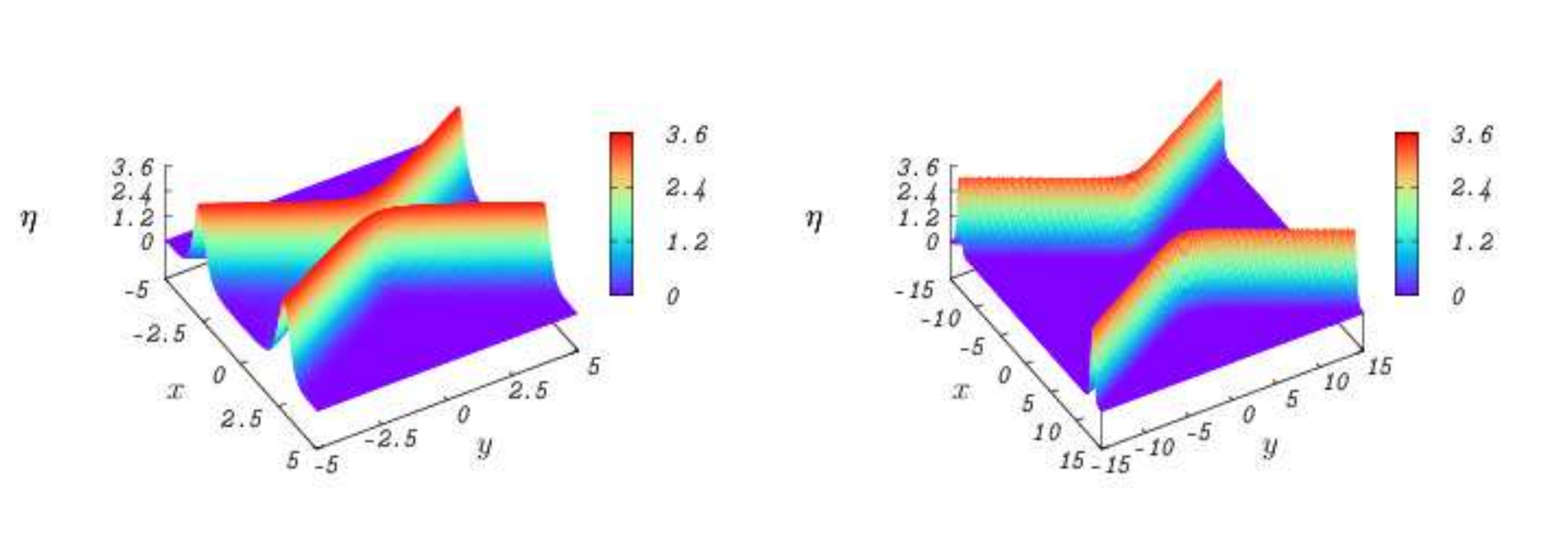}
\caption{Surface waves corresponding to the one-soliton solution of the ecKP-II equation (\ref{H1}) with $K=1.5, L=0$ for $t=0$ (left), $t=2$ (right).}\label{sw3}
\end{figure}

For
\begin{equation*}
\frac{6^{1/3}}{2\Delta}\leq K^2\leq \frac{6^{1/3}}{\Delta} (\approx3.63 \quad \mbox{for} \quad \Delta = \frac 12)
\end{equation*}
the condition (\ref{sin2b}) is satisfied for 
\begin{equation*}
\abs{\sin\beta}=\sqrt{2-\dfrac{2\Delta}{6^{1/3}}K^2-\frac{2t}{\gamma}\sech\alpha},
\end{equation*}
and the wave splits into two deformed line solitons (shown in Fig.~\ref{sw3}).

If
\begin{equation*}
K^2 > \frac{6^{1/3}}{\Delta}
\end{equation*}
then the argument of (\ref{H1}) cannot be equal to zero, and the wave continuously attenuates with time. (The large values of $K$ lie outside of the range of validity of the model.)

Similarly, we can consider the case $L\ne0$ (without loss of generality we assume that $L>0$, see (\ref{L})). If 
\begin{equation*}
\dfrac{\Delta}{6^{1/3}}\left[K^2+3L^2\right]>1, 
\end{equation*}
then the argument \eqref{arg} of $\sech^2$ is negative for $L \sin \beta < 0$ already at $t=0$, and it decreases with the increase of $t$. For $L\sin\beta=\abs{L}$ the argument is closer to zero than for other values of $\beta$. Therefore $\eta$ has a maximum for $L\sin\beta=\abs{L}$ as a function of $\beta$. 
However, as we mentioned above, such large values of $K$ and $L$ are likely to be outside of the range of validity of the model, we mention them here only  for the completeness of our analysis.

For sufficiently small $K$ and $L$ equation
 \eqref{sin2b} yields
\begin{equation*}
\sin\beta=\Delta^{1/2}6^{1/3}L\tanh\alpha\pm\sqrt{D},
\end{equation*}
where
\begin{equation*}
D=2-\dfrac{2\Delta}{6^{1/3}}K^2-\Delta6^{1/3}L^2\sech^2\alpha-\dfrac{2t}{\gamma}\sech\alpha.
\end{equation*}
Then, for $\alpha\to\infty$ we obtain
\begin{equation*}
\sin\beta=\Delta^{1/2}6^{1/3}L\pm\sqrt{2-\dfrac{2\Delta}{6^{1/3}}K^2}.
\end{equation*}

Therefore there exist several natural domains for  the values of the 
parameters $K$ and $L$. Let us note that if $2-\dfrac{2\Delta}{6^{1/3}}K^2<0$, then the solution (\ref{H1}) continuously attenuates, and we do not consider this range of values. Let $2-\dfrac{2\Delta}{6^{1/3}}K^2 > 0$. Then, there are several cases.

\noindent
(i) If $\abs{\Delta^{1/2}6^{1/3}L\pm\sqrt{2-\dfrac{2\Delta}{6^{1/3}}K^2}}>1$,   the solution (\ref{H1})  has the form of a deformed nearly-elliptic wave (shown in Fig.~\ref{sw1.5}). Indeed, in this case the argument of (\ref{H1}) can not be equal to zero for sufficiently large $\alpha$, and therefore the solution is localised.  For sufficiently small $L$ the above inequality implies:
\begin{eqnarray*}
&&\Delta^{1/2}6^{1/3}L+\sqrt{2-\dfrac{2\Delta}{6^{1/3}}K^2}>1 \quad \text{and}\\
&&\Delta^{1/2}6^{1/3}L-\sqrt{2-\dfrac{2\Delta}{6^{1/3}}K^2}<-1,
\end{eqnarray*}
which yields
\begin{equation*}
\sqrt{2-\dfrac{2\Delta}{6^{1/3}}K^2}>1+\Delta^{1/2}6^{1/3} L,
\end{equation*}
and
\begin{equation}
K^2+3L^2+\dfrac{6^{2/3}}{\Delta^{1/2}} L<\dfrac{6^{1/3}}{2\Delta}.
\label{ineq.KL}
\end{equation}
It is easy to see that if \eqref{ineq.KL} holds, then for any $\beta$
\begin{equation*}
2-\sin^2\beta-\dfrac{2\Delta}{6^{1/3}}\left[K^2+3L^2\right]>
2\Delta^{1/2}6^{1/3} L \abs{\sin\beta}.
\end{equation*}
Therefore we can introduce the notations
\begin{gather*}
2-\sin^2\beta-\dfrac{2\Delta}{6^{1/3}}\left[K^2+3L^2\right]=A(\beta)\cosh(\alpha_0),\\
2\Delta^{1/2}6^{1/3}L\sin\beta=A(\beta)\sinh(\alpha_0),
\end{gather*}
where
\begin{eqnarray*}
&&A(\beta)= \{ \left(2-\sin^2\beta-\dfrac{2\Delta}{6^{1/3}}\left[K^2+3L^2\right]\right)^2  \\
&& -4\Delta6^{2/3}L^2\sin^2\beta \}^{1/2},\\
&&\tanh\alpha_0=\dfrac{2\Delta^{1/2}6^{1/3}L\sin\beta} {2-\sin^2\beta-\dfrac{2\Delta}{6^{1/3}}\left[K^2+3L^2\right]}.
\end{eqnarray*}
In these notations equation \eqref{arg} takes the form
\begin{equation}
\cosh(\alpha+\alpha_0)=\dfrac{2t}{A(\beta)\gamma}.
\label{arg-2}
\end{equation}
Therefore in this case for any $\beta$ and for sufficiently large $t$ there exists the value of the parameter $\alpha$ such that the argument \eqref{arg} will be equal to zero. Because all functions are continuous and differentiable, the maximum of $H(\tau,\zeta,\nu)$ will be attained along some smooth closed curve. For sufficiently small $t$ and for any $\alpha,\beta$ the value of the function $H(\tau,\zeta,\nu)$ will be less than $K^2/2$. Since the sign of $\alpha_0$ depends on the sign of $\beta$, for $L\ne0$ it follows that the solution will be asymmetric with respect to $\beta$. For $L\sin\beta<0$ the wave will be wider than for $L\sin\beta>0$ (shown in Fig.~\ref{sw1.5}).

\noindent
(ii) If $\Delta^{1/2}6^{1/3}L+\sqrt{2-\dfrac{2\Delta}{6^{1/3}}K^2}>1$ and $1 > \Delta^{1/2}6^{1/3}L-\sqrt{2-\dfrac{2\Delta}{6^{1/3}}K^2}>-1$,  the nearly-elliptic wave breaks for negative~$y$, producing a single deformed line-soliton (shown in Fig~\ref{sw2}).

\begin{figure}[htb]
    \centering
\includegraphics[width=8.5cm]{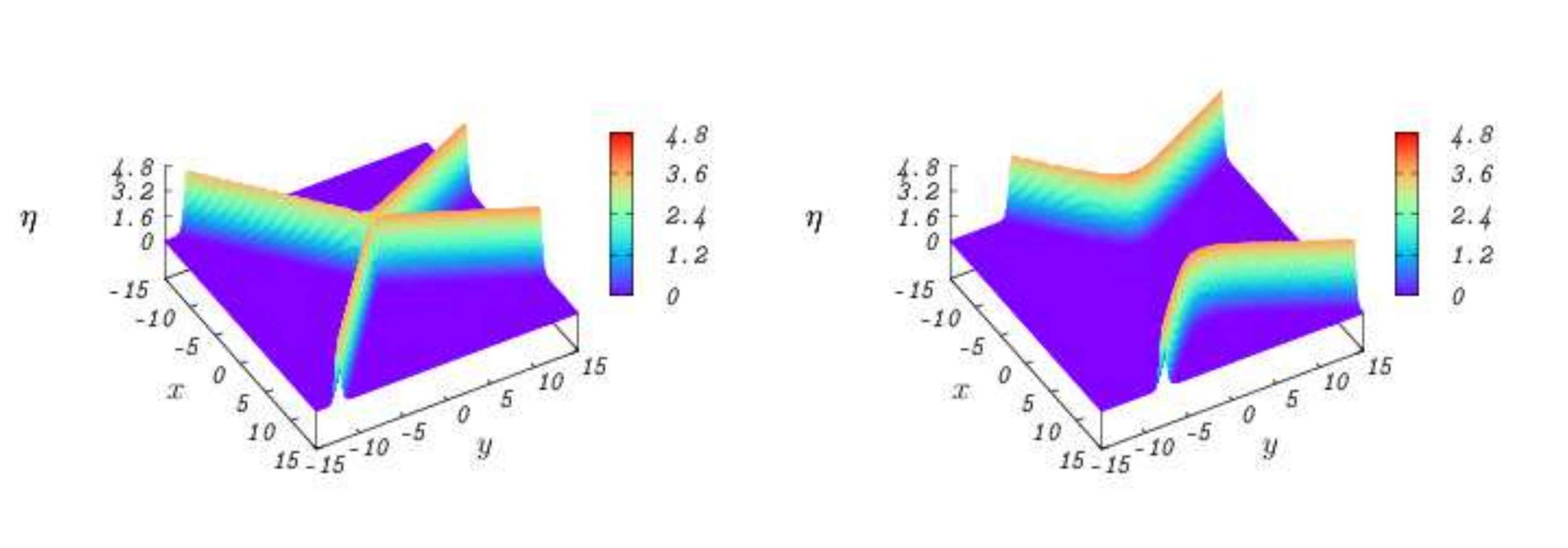}
\caption{Surface waves corresponding to the one-soliton solution of the ecKP-II equation (\ref{H1}) with $K=1.6, L=0.1$ for $t=0$ (left), $t=2$ (right).}\label{sw4}
\end{figure}

\noindent
(ii) If  $\abs{\Delta^{1/2}6^{1/3}L\pm\sqrt{2-\dfrac{2\Delta}{6^{1/3}}K^2}}<1$,  the nearly-elliptic wave breaks both for negative and positive~$y$, producing a pair of deformed line-solitons (shown in Fig.~\ref{sw4}).

\begin{figure}[htb]
    \centering
\includegraphics[width=8.5cm]{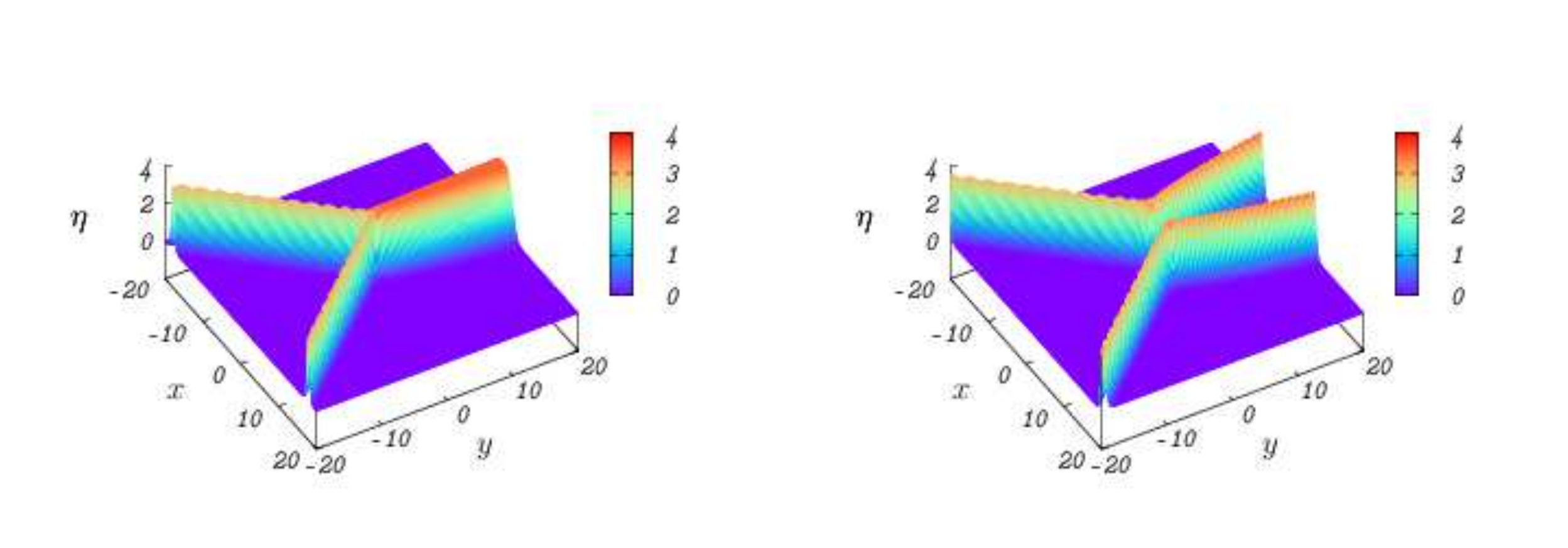}
\caption{Surface waves corresponding to the exceptional one-soliton solution of the ecKP-II equation (\ref{H1}) with $K=1.5$ and $L$ defined by (\ref{LY}) ($L \approx 0.1$)  for $t=0$ (left) and $t=1$ (right).}\label{Y}
\end{figure}

Apart from these generic cases, there are also some exceptional cases (corresponding to the boundaries between the generic cases). To illustrate that, we show a solution corresponding to the condition
\begin{equation}
\Delta^{1/2}6^{1/3}L+\sqrt{2-\dfrac{2\Delta}{6^{1/3}}K^2}=1,
\label{LY}
\end{equation}
which is the borderline case in between the last two generic cases. The solution is shown in Fig.~\ref{Y}, and it can be interpreted as describing the splitting of the wave looking like a resonant Y-soliton (or `Miles soliton' \cite{Miles}), known from the theory of the original KP equation, into two deformed line solitons. The existence of such solutions might indicate the instability of the Y-soliton with respect to perturbations in the area of the crossing.

\begin{figure}[htb]
    \centering
\includegraphics[width=8.5cm]{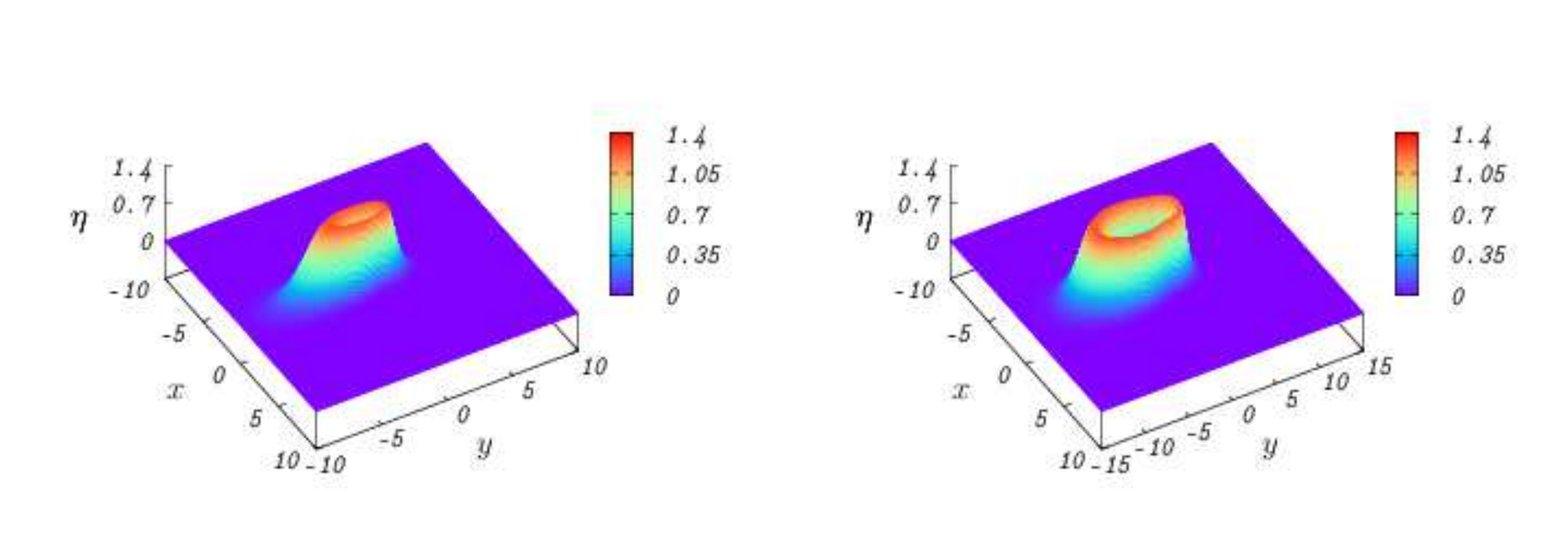}
\caption{Surface wave corresponding to the solution of the ecKP-II equation obtained from the canonical KP-soliton with $k_1=0.5, k_2=-0.4$ for $t=0.5$ (left), $t=1$ (right).}\label{sw5}
\end{figure}

Let us note that the one-soliton solution of the KP-II equation used to obtain the solution \eqref{H1} of the ecKP-II equation  above  can be written in the form of the `canonical' soliton for the KP-II equation
\begin{eqnarray*}
&&U(\tau,\xi,Y)=2\partial_{\xi}^2\ln(\psi_1+\psi_2), \quad \mbox{where}\\
&& \psi_j=\exp(k_j\xi+\sigma k_j^2Y-4k_j^3\tau),
\end{eqnarray*}
if $K=k_1-k_2$ and $L=(k_1+k_2)\sigma$ (we let $\delta_0 = 0$). Then, 
the conditions on $k_j$ follow from the conditions on $K$ and $L$, discussed above. We show in Fig.~\ref{sw5} the surface wave corresponding to the solution of ecKP-II equation obtained as the image of the canonical one-soliton solution of KP-II equation with $\sigma=1, k_1=0.5, k_2=-0.4$ under the map (\ref{eckpmap}).

\begin{figure}[htb]
    \centering
\includegraphics[width=8.5cm]{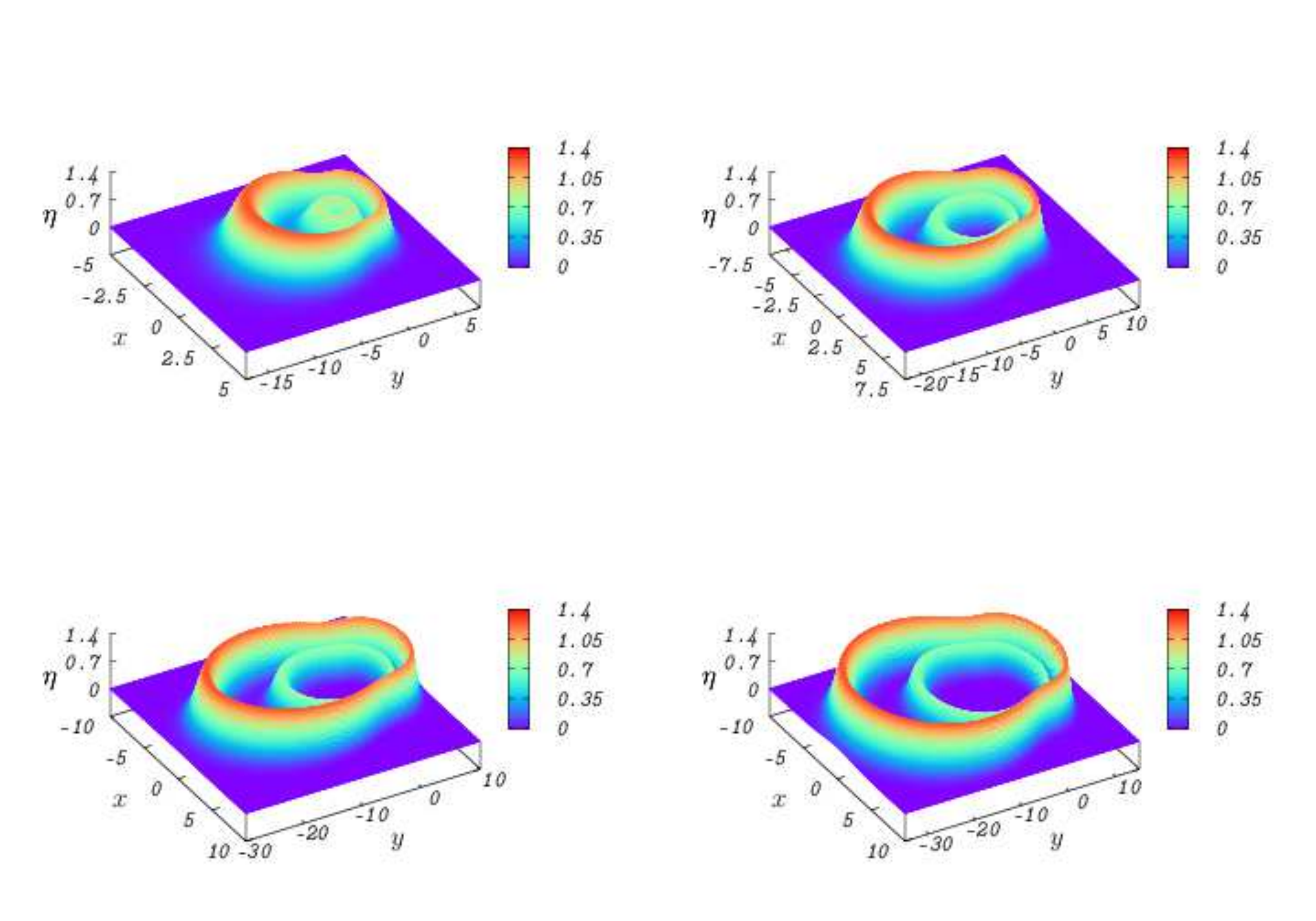}
\caption{Surface waves corresponding to the two-soliton solution of the ecKP-II equation with $k_1=0.5,k_2=-0.4$, $k_3=0.4,k_4=-0.3$ for $t=1$ (top left), $t=2$ (top right), $t = 3$ (bottom left), $t = 4$ (bottom right).}\label{sw6}
\end{figure}

\begin{figure}[htb]
    \centering
\includegraphics[width=8.5cm]{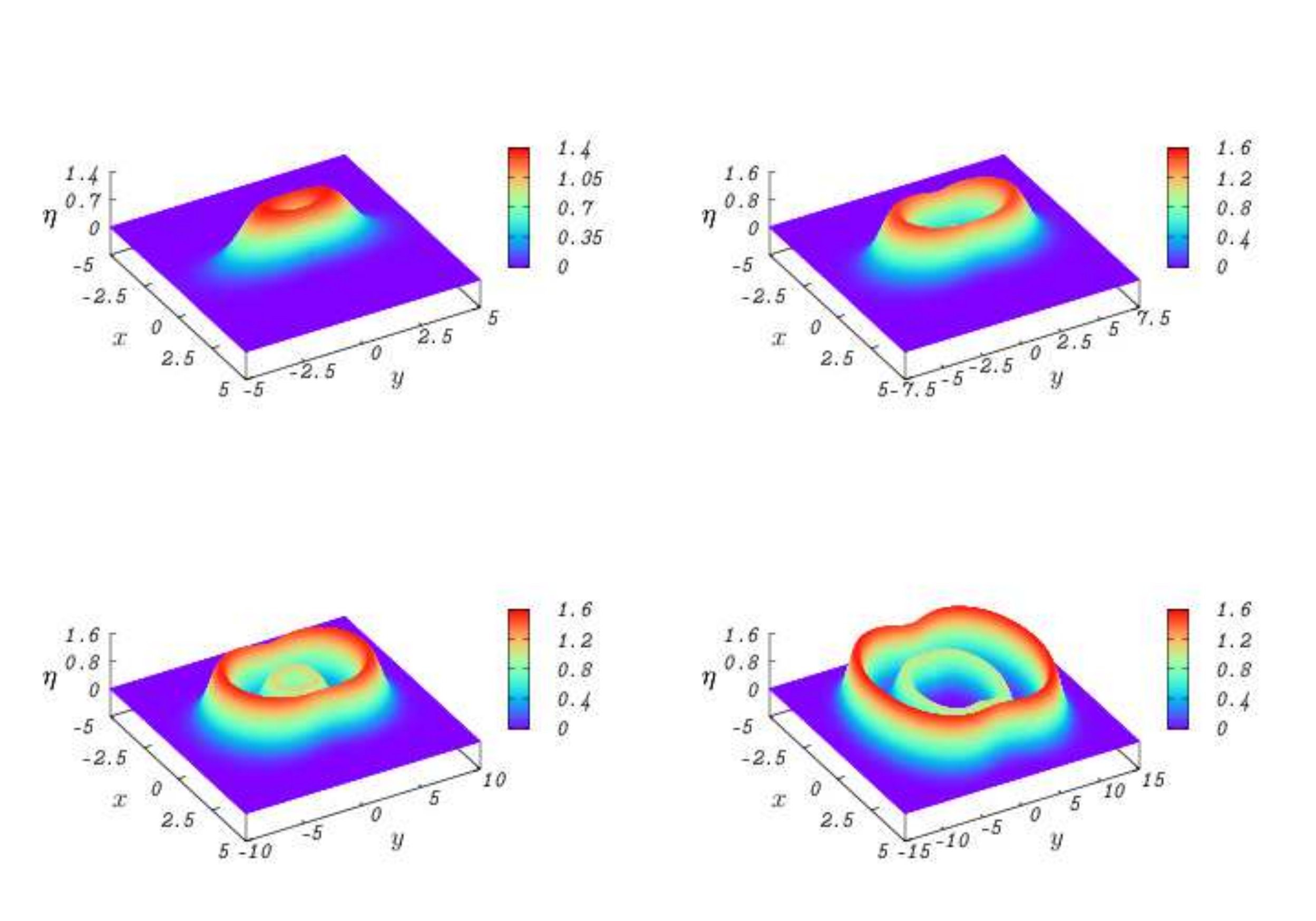}
\caption{Surface waves corresponding to the two-soliton solution of the ecKP-II equation with $k_1=0.5,k_2=-0.5$, $k_3=0.4,k_4=-0.4$ for $t=0$ (top left), $t=0.5$ (top right), $t = 1$ (bottom left), $t = 2$ (bottom right).}\label{sw7}
\end{figure}

Using Darboux transformations, one can obtain the canonical two-soliton solution of the KP-II equation in the form
\begin{eqnarray}
&&U(\tau,\xi,Y)=2\partial_{\xi}^2\ln(\phi_1\phi_{2\xi}-\phi_2\phi_{1\xi}), \quad \mbox{where} \nonumber \\
&&\phi_1=\psi_1-\psi_2,\quad \phi_2=\psi_3+\psi_4 \label{2sol}
\end{eqnarray}
(up to the phase shifts, which can be added to the phases).
Some particular surface waves corresponding to the two-soliton solutions of the ecKP-II equation (obtained as the image of  (\ref{2sol}) under the map (\ref{eckpmap}))  are shown in  Fig.~\ref{sw6} and Fig~\ref{sw7} (asymmetric and symmetric two-soliton nearly-elliptic waves, respectively).

Finally, let us choose $W_e=2/3$ and consider the ecKP-I lump solution (the image of the KP-I lump under the map (\ref{eckpmap}))
\begin{equation*}
H(\tau,\zeta,\nu)=\dfrac{4\kappa(1-\kappa(\zeta+\tau\nu^2/12-3\kappa\tau)^2 +\kappa^2(\tau^2-a^2)\nu^2)} {(1+\kappa(\zeta+\tau\nu^2/12-3\kappa\tau)^2+\kappa^2(\tau^2-a^2)\nu^2)^2}.
\end{equation*}
It is easy to see that for $\sin\beta=0$ the wave elevation $\eta$ has the form
\begin{equation*}
\eta=-\frac{4}{6^{1/3}}\sqrt{\frac{a}{\gamma}}f(\zeta-3\kappa\tau),
\end{equation*}
where the function
\begin{equation*}
f(X)=\dfrac{4\kappa(1-\kappa X^2)}{(1+\kappa X^2)^2}
\end{equation*}
has one high maximum for $X=0$ and two weak minima for $X=\pm\sqrt{3/\kappa}$, where
\begin{equation*}
f(0)=4\kappa, \quad f(\pm\sqrt{3/\kappa})=-\kappa/2.
\end{equation*}

Therefore, for sufficiently large values of $t$ the wave elevation $\eta$ has two deep minima
\begin{equation*}
\eta_{min}=-\dfrac{16\kappa}{6^{1/3}}\sqrt{\dfrac{a}{\gamma}}
\end{equation*}
for
\begin{equation*}
\sin\beta=0,\quad \cosh\alpha=\frac{t}{\gamma \left (1+\dfrac{3\kappa\Delta}{6^{1/3}}\right )}.
\end{equation*}

\begin{figure}[htb]
    \centering
\includegraphics[width=8.5cm]{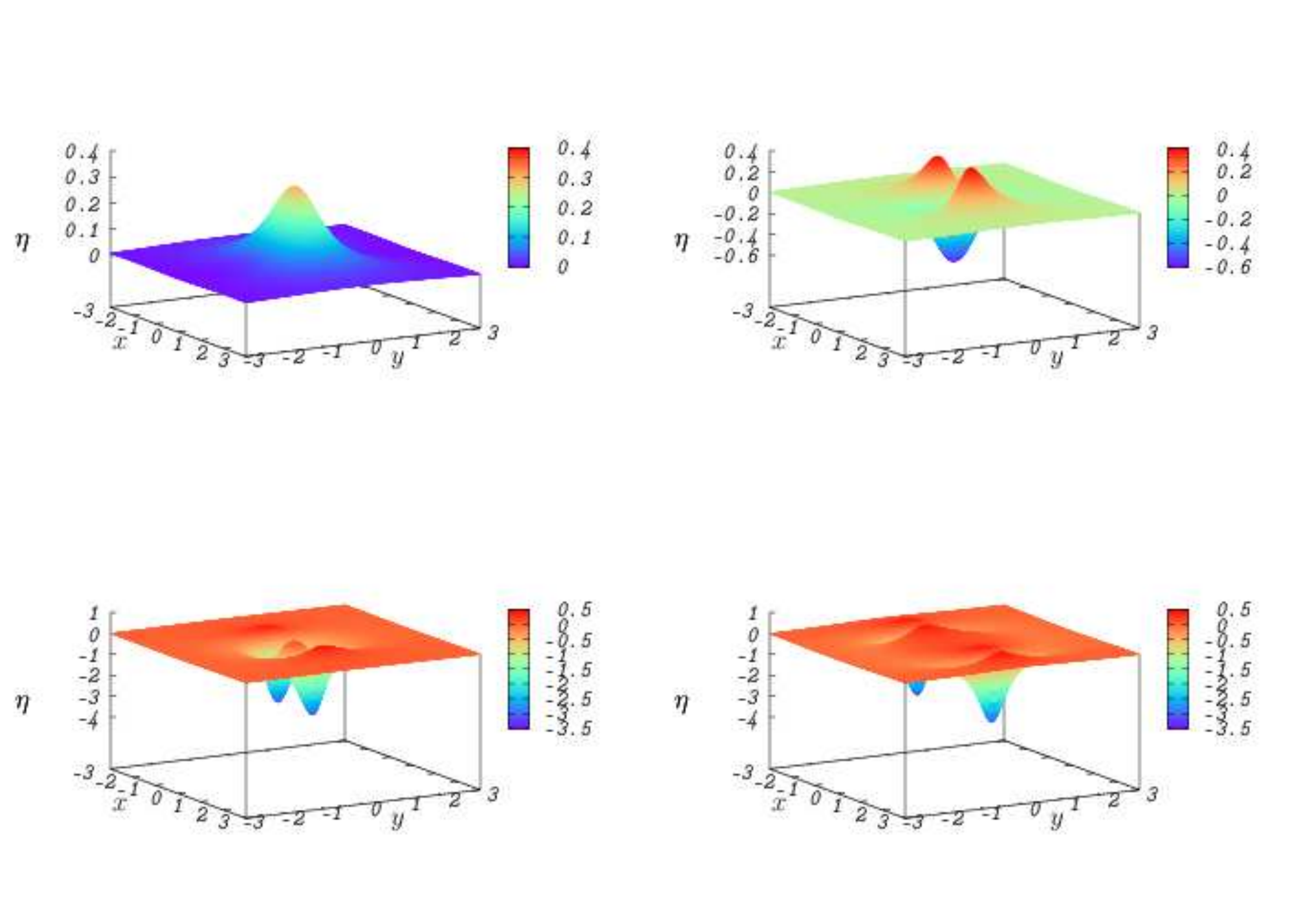}
\caption{Surface waves corresponding to the ecKP-I lump solution with $\kappa=0.25$, for $t=0$ (top left), $t=0.5$ (top right), $t = 1$ (bottom left), $t = 2$ (bottom right).}\label{sw8}
\end{figure}

The corresponding surface wave elevation $\eta$ is  plotted in Fig.~\ref{sw8} for $\gamma = 1, a=2, \Delta = 1/2, W_e=2/3$ and $\kappa=0.25$ .

\section{Concluding remarks}

In this paper we have derived and studied a new integrable version of the Kadomtsev-Petviashvili equation associated with the elliptic-cylindrical geometry of the wave fronts. The derivation was given in the context of surface gravity waves, but the equation can be readily derived in other physical contexts. We found transformations linking the derived model with the two classical versions of the KP equation, associated with the Cartesian and cylindrical geometries of the wave fronts, and the Lax pair for the new equation. We also completely  classified approximate solutions for the surface gravity waves corresponding to the one-soliton solution of the ecKP equation, as well as discussing some other solutions.

In our derivation a large distance variable has been used in preference to large time, although one can also use the large time variable. The dimensional form of the derived equation is given by
\begin{eqnarray*}
&&\left [ 2 \left (\eta_t + \frac cd \eta_\phi\right )  - \frac{3}{h_0} \eta \eta_t - \frac{h_0^2}{c^2} \left ( \frac 13 - W_e \right ) \eta_{ttt}  \right. \\
&&\left . + \frac cd \frac{\phi}{\phi^2-1} \eta + \frac{\psi^2}{\phi^2-1} \eta_t \right ]_t - \frac{c^2}{d^2} \frac{1}{\phi^2-1} \eta_{\psi \psi} = 0,
\end{eqnarray*}
where $\eta$ is the free surface elevation, $t$ is time, $\phi = \cosh \alpha$ and $\psi = \sin \beta$ are variables related to the elliptic cylindrical coordinates, $h_0$ is the unperturbed fluid depth, $c = \sqrt{g h_0}$ is the linear long-wave speed,  $d$ is half of the distance between the foci of the coordinate lines (say, the boundary of the wave source), and $W_e$ is the Weber number.
The key non-dimensional parameters used in the paper are expressed via the dimensional parameters as follows
$$
A = \frac{d \lambda^3 h_s^6}{h_0^{10}}, \quad \gamma = \frac d \lambda, \quad \Delta = \frac{\lambda^2 h_s^4}{h_0^6},
$$
where $\lambda$ is the wave length, while $\delta = \frac{h_0}{\lambda}$ and $ \epsilon = \frac{h_s}{h_0}$. To derive the ecKP equation, we required that $\Delta$ is a small parameter. We also note that for any given values of $A, \gamma$ and $\Delta$ there exists a range of the physical validity of the model, as can be seen from the expressions above.

The importance of the model to particular applications has not been discussed in this paper, and it is an open question at the moment. Another open question is the study of the wave instabilities within the framework of the ecKP equation, continuing the lines of research for the KP equation \cite{Zak} and the cKP equation \cite{Ostr}.

In our paper we considered only some simple solutions of the derived equation. Recently, there has been significant progress in the classification of soliton solutions of the KP equation with applications to water wave problems (see \cite{BC, B, CK, YLK} and references therein). It would be interesting to see the counterpart of this classification for the derived equation, and for the approximate solutions for surface waves.

The derivation of the ecKP equation from the full set of Euler equations opens the way to the study of  internal and surface waves on a current for a stratified fluid, as well as accounting for the effects of variable background and Earth's rotation, which will be reported elsewhere. It paves the way for other applications, for example, in the context of matter waves in Bose-Einstein condensates (e.g., \cite{Huang, TDP}), since the hydrodynamic form of the Gross-Pitaevskii equation is similar to the problem formulation (\ref{Euler}). Also, recent studies of `spherical nebulons' \cite{GT2}, based on the spherical KP equation, can be extended since the ellipsoidal KP equation, associated with the ellipsoidal coordinates, can be derived from the equations for `a dusty plasma' along the lines discussed in this paper.

Finally, it is natural to ask a question whether one can derive other 
versions of the KP equation, associated with other coordinate systems (i.e.\ with other wave geometries), and whether one can find the general description of all admissible maps of the type discussed in section 3, associated with the problem formulation (\ref{Euler}).

\section{Acknowledgments}

We thank G.A. El, E.V. Ferapontov and R.H.J. Grimshaw for useful discussions, and referees for constructive comments and helpful references.
KRK and AOS acknowledge support and hospitality  of the Institut de Math\'ematiques de 
Bourgogne, where they held visiting positions in the spring-summer of 2012,
which has made this collaboration possible. CK and VBM  thank for financial support by  the ANR via the program  ANR-09-BLAN-0117-01. 

\section{Appendix A}

In the variables
\begin{eqnarray*}
&&\zeta = \frac{\epsilon^2}{\delta^2} \left ( \gamma \cosh \alpha - t \right ),  \\
&&R = \frac{\epsilon^6}{\delta^4} \gamma \cosh \alpha, \quad \nu =  \frac{\delta}{\epsilon^2} \sin \beta, \\
&& u =  \frac{\epsilon^3}{\delta^2} U, \quad   v =  \frac{\epsilon^5}{\delta^3}  V, \quad w = \frac{\epsilon^5}{\delta^4}  W, \\
&&\eta = \frac{\epsilon^3}{\delta^2} H, \quad p = \frac{\epsilon^3}{\delta^2} P,
\end{eqnarray*}
 the problem formulation (\ref{1}) - (\ref{7}) assumes the form 
{\small
\begin{eqnarray*}
&&- U_\zeta  + P_\zeta   + \Delta \left [UU_\zeta + W U_z + P_R  \right. \nonumber  \\
&& \left. - \frac{R - \sqrt{R^2-A^2} }{R^2 - A^2} \left (\nu P_{\nu} + R  \nu^2 P_\zeta \right )  \right ]  + O(\Delta^2) = 0, \label{11} \\[2ex]
&&-V_\zeta  + \frac{1}{\sqrt{R^2-A^2}} P_\nu  + \frac{R - \sqrt{R^2-A^2} }{\sqrt{R^2 - A^2}} \nu P_\zeta  \nonumber \\
&& + \Delta \left [ U V_{\zeta} + W V_z   + \frac{R-\sqrt{R^2-A^2}}{\sqrt{R^2-A^2}} \nu P_{R} \right.  \nonumber \\
&& \left. +  \frac{(R-\sqrt{R^2-A^2}) (R^2+A^2)}{2 (R^2-A^2)^{3/2}} \nu^3 P_{\zeta} \right.  \nonumber \\
&& \left .  - \frac{(R-\sqrt{R^2-A^2})^2 + R^2}{2 (R^2-A^2)^{3/2}} \nu^2  P_{\nu} \right ]
+ O(\Delta^2) = 0, \label{12} \\[2ex]
&&P_z - \Delta W_\zeta + O(\Delta^2) = 0, \label{13} \\[2ex]
&&U_\zeta +  W_z + \Delta \left [ U_R - \frac{R - \sqrt{R^2-A^2}}{R^2-A^2} \left (R \nu^2 U_\zeta+  \nu U_\nu   \right ) \right . \nonumber \\
&& + \left .  \frac{1}{\sqrt{R^2 - A^2}} \left ( (R - \sqrt{R^2-A^2}) \nu V_{\zeta} +  V_\nu +  U \right )  \right ] \nonumber \\
&&+ O(\Delta^2) = 0, \label{14} 
\end{eqnarray*}
\begin{eqnarray*}
&&P|_{z = 1 + \Delta H(\zeta, R, \mu)} = H - \Delta W_e \ H_{\zeta \zeta} + O(\Delta^2), \label{15} \\[2ex]
&&W|_{z = 1 + \Delta H(\zeta, R, \mu)} = -H_\zeta + \Delta U H_\zeta + O(\Delta^2), \label{16} \\[2ex]
&&W|_{z=0} = 0, \label{17}
\end{eqnarray*}
}
where
$
\Delta = \frac{\epsilon^4}{\delta^2}, \quad A = \gamma \frac{\epsilon^6}{\delta^4}.
$
Here, we have not shown the explicit form of the higher-order terms in the small parameter $\Delta$ (denoted by $O(\Delta^2)$) since these terms are not needed in the derivation of our asymptotic equation.

\section{Appendix B}

Quasi-periodic (multiphase) solutions to the KP equation can be given in terms of
multi-dimensio\-nal theta functions on compact Riemann
surfaces of arbitrary genus $n$ (see  \cite{K}) in the form 
\begin{equation}
U(\xi, Y, \tau) = 2 \partial_{x}^{2}\ln \Theta [\xi p + Y v + \tau q + l] + C,
\label{theta}
\end{equation} 
where $\Theta$ is the Riemann theta function, $p, v, q, l$ are 
periods of certain integrals 
on this surface, and 
$C$ is constant with respect to the coordinates $\xi$, $Y$ and 
$\tau$, see \cite{BBEIM,FK2} for details. For a given Riemann surface 
and a given point on it, these 
quantities are uniquely determined.

\begin{figure}[h]
    \centering
\includegraphics[width=8.5cm]{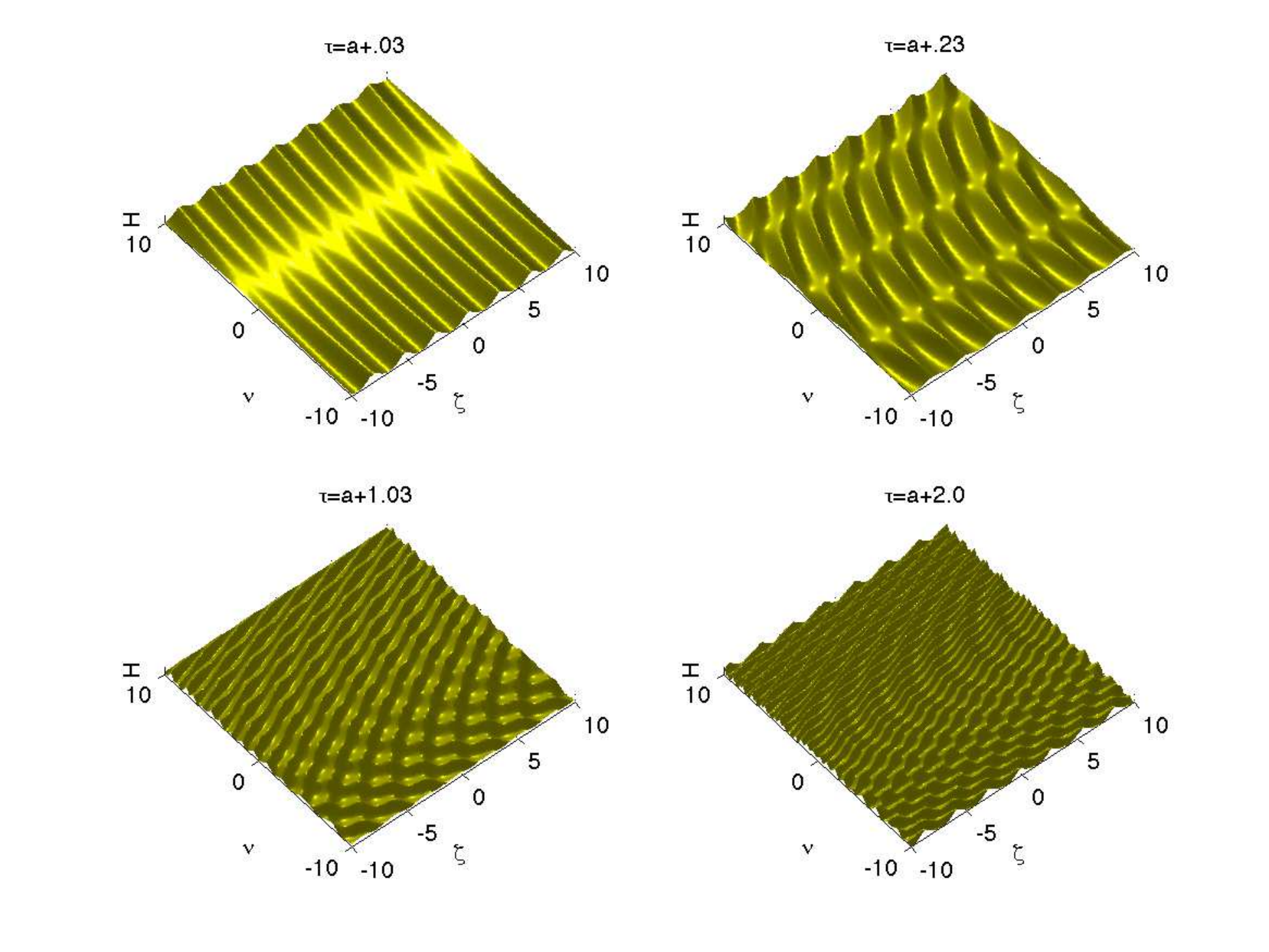}
\caption{Genus $2$ solution (\ref{theta}) to the ecKP-II equation for $a=0.01$ generated by
the curve $w^{2}=\prod_{i=1}^{6}(z-e_{i})$,
$e_{1}=-3,e_{2}=-2,e_{3}=-1, e_{4}=0,e_{5}=1,e_{6}=2$ for several
values of $\tau$. 
}\label{12a}
\end{figure}

In genus 2 all such surfaces are hyperelliptic.
In this case we consider the
hyperelliptic curve with branch points $-1,-2,-3,0,1,2$. These
solutions are  numerically evaluated with the spectral code by
Frauendiener and Klein
\cite{FK1,FK2}. The
related solutions to
the ecKP-II equation are generated from the corresponding  solutions of the
KP-II equation via the map (\ref{eckpmap}) with $a=0.01$. We clearly
see in Fig.~\ref{12a} the
formation of intersecting families of parabolic fronts.

\begin{figure}[h]
   \centering
\includegraphics[width=8.5cm]{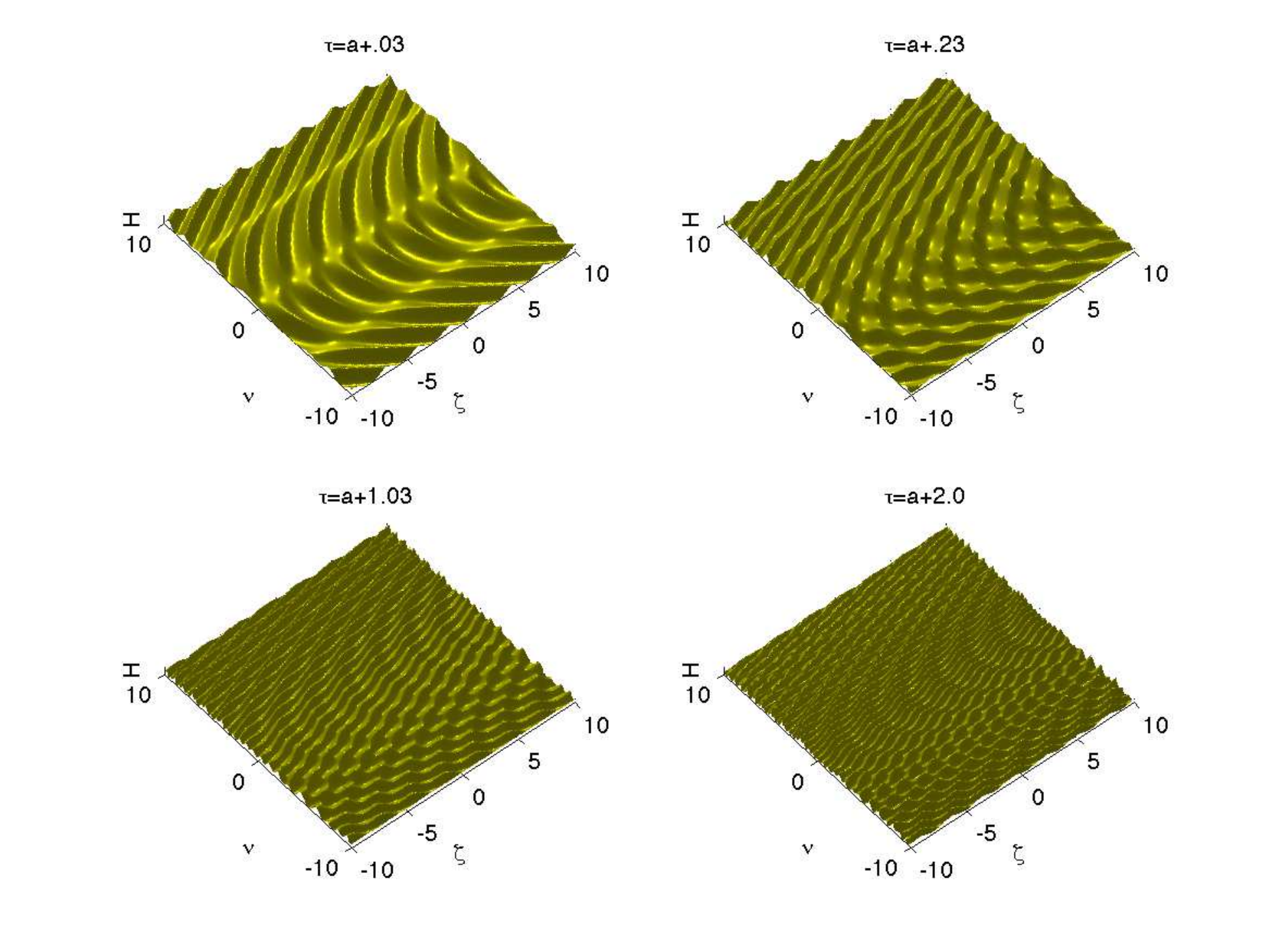}
\caption{Genus $2$ solution (\ref{theta}) to the ecKP-II equation for $a=1$ generated by
the curve $w^{2}=\prod_{i=1}^{6}(z-e_{i})$,
$e_{1}=-3,e_{2}=-2,e_{3}=-1, e_{4}=0,e_{5}=1,e_{6}=2$ for several
values of $\tau$. 
}
\label{eckp2}
\end{figure}

In the same setting with $a=1$, i.e., a theta-functional solution to
the ecKP-II equation, the formation of curved profiles is already
present for small $\tau-a$ as can be seen in Fig.~\ref{eckp2}. Both
cases asymptotically coincide for $\tau\to\infty$.

\begin{figure}[h]
   \centering
\includegraphics[width= 8.5cm, height=3cm
]{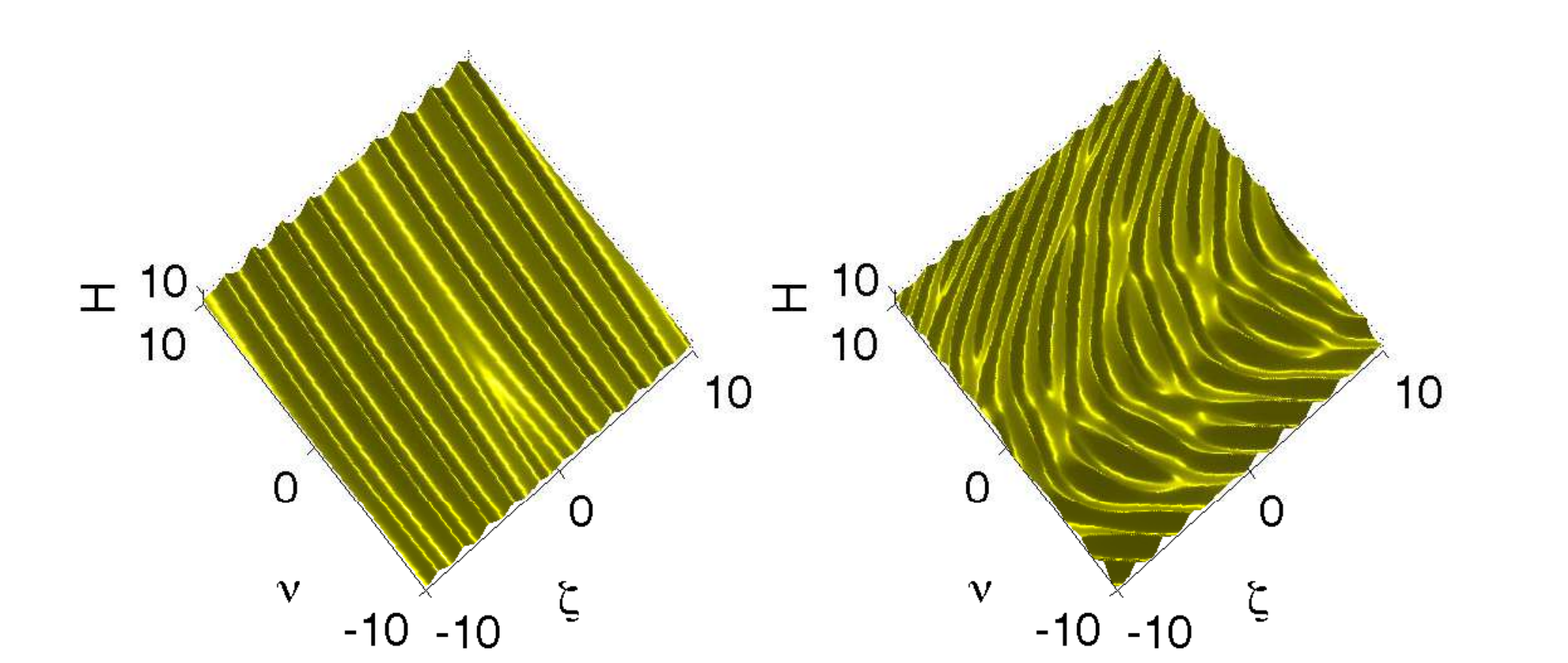}
\caption{Genus $3$ solution (\ref{theta}) to the ecKP-II equation for $a=0.01$ (left) and $a=1$ (right), generated by
the curve $w^{2}=\prod_{i=1}^{8}(z-e_{i})$,
$e_{1}=-5,e_{2}=-4,e_{3}=-3, e_{4}=-2,e_{5}=-1,e_{6}=0, e_{7}=1,
e_{8}=2$ at $\tau=a+0.01$. 
}\label{eckp3}
\end{figure}

In higher genus, the solutions are $g$-phase solutions, i.e., they
have more structure as can be seen in Fig.~\ref{eckp3}. We consider 
here again hyperelliptic surfaces. The close to cKP
solutions are for small time essentially independent of the transversal coordinate.

\end{document}